\documentclass[8.5pt,twoside,twocolumn]{article}
\usepackage[super,sort&compress,comma]{natbib} 
\usepackage{mhchem}
 \usepackage{times,subfigure}
\usepackage{amssymb}
\usepackage{amsmath}

\usepackage{sectsty}
\usepackage{balance} 

\usepackage{graphicx}
\usepackage{lastpage}
\usepackage[format=plain,justification=raggedright,singlelinecheck=false,font=small,labelfont=bf,labelsep=space]{caption} 
\usepackage{fancyhdr}
\usepackage{color}
\definecolor{grey}{rgb}{0.5,0.5,0.5}

\begin{document}

\thispagestyle{plain}
\fancypagestyle{plain}{
\fancyhead[L]{\includegraphics[height=8pt]{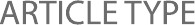}}
\fancyhead[C]{\hspace{-1cm}\includegraphics[height=20pt]{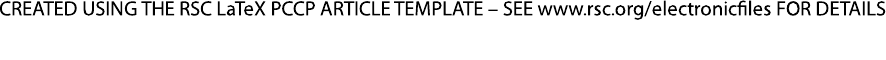}}
\fancyhead[R]{\includegraphics[height=10pt]{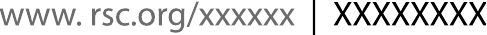}\vspace{-0.2cm}}
\renewcommand{\headrulewidth}{1pt}}
\renewcommand{\thefootnote}{\fnsymbol{footnote}}
\renewcommand\footnoterule{\vspace*{1pt}%
\hrule width 3.4in height 0.4pt \vspace*{5pt}} 
\setcounter{secnumdepth}{5}

\makeatletter 
\def\subsubsection{\@startsection{subsubsection}{3}{10pt}{-1.25ex plus -1ex minus -.1ex}{0ex plus 0ex}{\normalsize\bf}} 
\def\paragraph{\@startsection{paragraph}{4}{10pt}{-1.25ex plus -1ex minus -.1ex}{0ex plus 0ex}{\normalsize\textit}} 
\renewcommand\@biblabel[1]{#1}            
\renewcommand\@makefntext[1]%
{\noindent\makebox[0pt][r]{\@thefnmark\,}#1}
\makeatother 
\renewcommand{\figurename}{\small{Fig.}~}
\sectionfont{\large}
\subsectionfont{\normalsize} 

\fancyfoot{}
\fancyfoot[LO,RE]{\vspace{-7pt}\includegraphics[height=9pt]{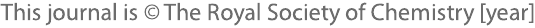}}
\fancyfoot[CO]{\vspace{-7.2pt}\hspace{12.2cm}\includegraphics{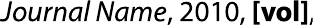}}
\fancyfoot[CE]{\vspace{-7.5pt}\hspace{-13.5cm}\includegraphics{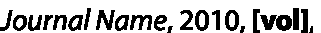}}
\fancyfoot[RO]{\footnotesize{\sffamily{1--\pageref{LastPage} ~\textbar  \hspace{2pt}\thepage}}}
\fancyfoot[LE]{\footnotesize{\sffamily{\thepage~\textbar\hspace{3.45cm} 1--\pageref{LastPage}}}}
\fancyhead{}
\renewcommand{\headrulewidth}{1pt} 
\renewcommand{\footrulewidth}{1pt}
\setlength{\arrayrulewidth}{1pt}
\setlength{\columnsep}{6.5mm}
\setlength\bibsep{1pt}

\newcommand{\dd}{\ensuremath{\mathrm{d}}}
\newcommand{\tu}{\ensuremath{\tilde{u}}}
\newcommand{\tz}{\ensuremath{\tilde{z}}}
\newcommand{\tG}{\ensuremath{\tilde{G}}}
\newcommand{\ta}{\ensuremath{\tilde{\alpha}}}
\newcommand{\tm}{\ensuremath{\tilde{\mu}}}
\newcommand{\tEr}{\ensuremath{\widetilde{Er}}}
\newcommand{\tDg}{\ensuremath{\widetilde{Dg}}}
\newcommand{\tDr}{\ensuremath{\widetilde{Dr}}}
\newcommand{\tAn}{\ensuremath{\widetilde{An}}}
\newcommand{\sgn}{\ensuremath{\mathrm{sgn}}}
\newcommand{\thalign}{\ensuremath{\theta_{\text{Leslie}}}}
\newcommand{\thsub}{\ensuremath{\theta_{\text{sub}}}}
\newcommand{\thturn}{\ensuremath{\theta_{\text{turn}}}}
\newcommand{\epsub}{\ensuremath{\epsilon_{\text{sub}}}}
\newcommand{\epmid}{\ensuremath{\epsilon_{\text{mid}}}}

\newcommand{\added}[1]{\textcolor{blue}{#1}}
\newcommand{\changed}[1]{\textcolor{red}{#1}}
\newcommand{\removed}[1]{\textcolor{grey}{#1}}

\twocolumn[
  \begin{@twocolumnfalse}
\noindent\LARGE{\textbf{The effect of anchoring on nematic flow in channels}}
\vspace{0.6cm}

\noindent\large{\textbf{Vera M. O. Batista$^{\ast}$, Matthew L. Blow and
Margarida M. Telo da Gama}}\vspace{0.5cm}

\noindent\textit{\small{\textbf{Received Xth XXXXXXXXXX 20XX, Accepted Xth XXXXXXXXX 20XX\newline
First published on the web Xth XXXXXXXXXX 200X}}}

\noindent \textbf{\small{DOI: 10.1039/b000000x}}
\vspace{0.6cm}

\noindent \normalsize{Understanding the flow of liquid crystals in microfluidic environments plays an important role in many fields, including device design and microbiology. 
We perform hybrid lattice-Boltzmann simulations of a nematic liquid crystal flowing under an applied pressure gradient in two-dimensional channels with various 
anchoring boundary conditions at the substrate walls. We investigate the relation between flow rate and pressure gradient and the corresponding profile of the nematic director, 
and find significant departures from the linear Poiseuille relation. We also identify a morphological transition in the director profile and explain this in terms of an instability 
in the dynamical equations. We examine the qualitative and quantitative effects of changing the type and strength of the anchoring. Understanding such effects may provide a useful 
means of quantifying the anchoring of a substrate by measuring its flow properties.}
\vspace{0.5cm}
 \end{@twocolumnfalse}
  ]



\footnotetext{\textit{$^{\ast}$E-mail: vmbatista@fc.ul.pt}}
\footnotetext{\textit{~Centro de F\'{i}sica Te\'{o}rica e Computacional, Faculdade de Ci\^{e}ncias da Universidade de Lisboa,Campo Grande, Ed-C8, P-1749-016 Lisboa, Portugal}}
\footnotetext{\textit{~Departamento de F\'{i}sica, Faculdade de Ci\^{e}ncias da Universidade de Lisboa, P-1749-016 Lisboa, Portugal}}



\section{Introduction}
\label{sec:intro}
Microfluidics is a major field of scientific research and technological innovation, exploited in ink-jet printing, lab-on-a-chip devices for chemical analysis, and 
smart wetting surfaces, among many other applications. A key topic is the pressure-driven flow of fluids within micron-scale channels~\citep{whitesides07}. In addition 
to simple fluids, there is considerable interest in the microfluidics of liquid crystals. Molecular liquid crystals are widely used in display devices due to their 
optical properties, and understanding the switching dynamics of such devices is essential for optimising their speed and efficiency~\citep{mcintosh00}. More recently, 
it has been appreciated that many biophysical systems - including microtubule bundles~\citep{sanchez12}, actin filaments~\citep{chakrabarti07}, and dense suspensions 
of microswimmers~\citep{baskaran09,yeomans12} - also have liquid crystalline properties, but at the colloidal instead of molecular level. The confined flows of these materials, 
often driven by their own activity, underpin many transport and motility processes in microbiology.

Furthermore, with regard to fluid transport and flow, liquid crystals offer functionality not achievable with simple fluids. When in contact with structured surfaces, 
the interplay between bulk effects such as elasticity and surface effects such as anchoring leads to rich behaviour including complex wetting 
transitions~\citep{patricio11,silvestre12} and the stabilisation of topological defects~\citep{dammone12,shams14}. Such intricate surface effects can 
be exploited in novel microfluidic applications. For example, confinement of a liquid crystal in a channel can lead to the formation of topological line defects, 
which may be utilised as rails for the controlled transport of colloids or droplets of a secondary fluid~\citep{sengupta13SM}.

Poiseuille's law, which applies to the laminar flow of an incompressible Newtonian fluid in a channel, predicts a linear relation between 
the rate of flow and the pressure difference applied, in analogy to the relation between electrical current and potential difference stated by Ohm's law. 
In the case of a non-Newtonian fluid, such as a liquid crystal, departures from this linear relation may be observed. We concentrate specifically on nematic 
liquid crystals. These are composed of rodlike molecules that possess no positional order in their arrangement, but do order in their orientation along a 
common axis called the director. Nematics exhibit rich hydrodynamics owing to the coupling between fluid motion and director orientation - a phenomenon termed backflow. 
In the presence of a velocity gradient, backflow leads to distinctive behaviours, namely either the perpetual rotation of the director (tumbling regime), or a 
steady state in which the director has a tendency to adopt a given angle relative to the velocity gradient (aligning regime)~\citep{leslie68}. Which regime occurs 
depends on the material properties of the liquid crystal. In this paper we concentrate on aligning liquid crystals. Backflow effects may be interpreted as an anisotropic 
viscosity, additional to the standard Newtonian viscosity. This dependence of viscosity upon director orientation was first observed by \citet{miesowicz}.

There have been a number of quantitative studies into Poiseuille flow of liquid crystals in channels or between parallel plates. 
Early calculations were carried out by \citet{ericksen61} and \citet{leslie66}. \citet{fishers69} carried out experiments of flow in a cylindrical tube, 
confirming a non-linear relation between flow rate and pressure gradient that depends on anchoring orientation. \citet{denniston00,denniston01CTPS} performed 
lattice Boltzmann simulations for the case of strong homeotropic anchoring, and reported a topological transition in the texture of the nematic as driving pressure is increased, 
which was confirmed in experiments and numerical calculations by~\citet{jewell09}. \citet{zhou07} performed calculations for two-dimensional flow in the 
low-flow-rate, strong-anchoring limit for the cases of homeotropic, planar and tilted anchoring. Sengupta \textit{et al.} \cite{sengupta13PRL,sengupta13IJMS} performed experiments and 
lattice Boltzmann simulations of flow in rectangular channels over a wider range of flow rates. \citet{feng99} and \citet{quintansCarou06} investigated, 
by analytical and numerical techniques, nematic flow between plates of narrowing or widening separation, while \citet{manneville76} and \citet{tarasov10} 
investigated the onset of instabilities in channel flow.

We are not aware of any study that considers the effect of the {\em strength} of the anchoring on the relation between driving pressure gradient and 
mass flow rate. 
Anchoring strength is an important factor in a wide range of topics including display device switching \cite{willman07}, 
wetting transitions on structured surfaces \cite{silvestre12} and in the onset of spontaneous flow in active systems \cite{voituriez05}. 
From an experimental point of view, the anchoring strength of a substrate is difficult to measure, and most methods involve the optical properties of the 
liquid crystal~\citep{barbero84,choi13}, or its response to applied electric~\citep{yokoyama85,nastishin99} or magnetic~\citep{andrienko98} fields. 
In this study we show that it is not only the anchoring type but also that the anchoring strength that has an important effect on microfluidic flows of nematics, 
which consequently has the potential to be used to quantify anchoring. In other words, a quantitative understanding of the effect 
of anchoring strength on flow may provide a means to measure the anchoring strength. 
Such an approach would be especially useful for colloidal liquid crystals, which do not exhibit an electromagnetic response.

In this paper, we perform lattice Boltzmann simulations of driven flow between two parallel plates, driven by a specified pressure gradient, 
and measure the flow rate versus pressure gradient. We confirm the transition reported by \citet{denniston01CTPS} and \citet{jewell09}, and find 
that it is driven by a dynamical instability, rather than free energy considerations. We also observe a very strong departure from the Poiseuille relation 
that does not relate to any obvious morphological transition. We check our results against calculations in the low- and high-flow limits, and find that in the 
low-flow limit, anchoring strength influences flow rate via a term that is cubic in the pressure gradient.

The paper is organised as follows. In section~\ref{sec:model} we describe the model we use to simulate the system. In section~\ref{sec:results} we present and analyse our results. 
Our principal focus, in section~\ref{sec:homeotropic}, is on the case where the anchoring at the walls of the channel is homeotropic (i.e. the preferred orientation of the 
director is perpendicular to the walls). In section~\ref{sec:planar} we present results for the case where the anchoring is non-degenerate planar (the director 
preferentially lies in a specified direction in the plane of the substrate - in this case the direction along the channel) 
and we do likewise for a channel where one wall has homeotropic and the other planar anchoring (which we call hybrid anchoring). We discuss these results 
and how they compare to the homeotropic case. We conclude in section~\ref{sec:conclusions}.

\section{The model}
\label{sec:model}

The nematic order of the fluid is expressed using a traceless, symmetric, tensorial order parameter~\citep{degennes93} called the Q-tensor,
\begin{equation}
Q_{\alpha\beta}=\tfrac{1}{2}S\left(3n_{\alpha}n_{\beta}-\delta_{\alpha\beta}\right)+\tfrac{1}{2}B\left(m_{\alpha}m_{\beta}-l_{\alpha}l_{\beta}\right),   \label{eqn:QTensor}
\end{equation}

\noindent where $\mathbf{n}$ is a unit vector denoting the director, and $S$ is the degree of nematic ordering. We note that the Q-tensor is invariant 
under $\mathbf{n}\rightarrow-\mathbf{n}$, reflecting the head-tail symmetry of the molecular ordering. In some circumstances there may be biaxial ordering of 
degree $B$, with $\mathbf{m}$ and $\mathbf{l}$ forming an orthonormal set with $\mathbf{n}$.

The free energy of the system is given by the functional over the fluid region $\mathcal{R}$ and the substrate walls $\mathcal{W}$
\begin{multline}
\mathcal{F}=\int_{\mathcal{R}}\Big\{\tfrac{2}{3}A\left(\tau^{*}S_{\text{nem}}^{-2}Q_{\alpha\beta}Q_{\beta\alpha} \right. \\
\left. -\tfrac{4}{3}(2+\tau^{*})S_{\text{nem}}^{-3}Q_{\alpha\beta}Q_{\beta\gamma}Q_{\gamma\alpha} \right.
\left. +\tfrac{2}{3}S_{\text{nem}}^{-4}\left[Q_{\alpha\beta}Q_{\beta\alpha}\right]^{2}\right) \\
\left.  + \tfrac{1}{2}L\partial_{\gamma}Q_{\alpha\beta}\partial_{\gamma}Q_{\alpha\beta}\Big\}\dd V + 
\int_{\mathcal{W}}\tfrac{1}{2}\alpha(Q_{\alpha\beta}^{\text{pref}}-Q_{\alpha\beta})^{2}\dd S, \right.   
\label{eqn:freeEnergy}
\end{multline}
\noindent where $A$, $L$ and $\alpha$ are positive coefficients for bulk, elastic and anchoring free energies respectively. $\tau^{*}$ is a reduced temperature, 
such that the nematic phase with $S=S_{\text{nem}}$ is favoured for $\tau^{*}<1$, and the isotropic (unordered) phase with $S=0$ is favoured 
when $\tau^{*}>1$. $Q_{\alpha\beta}^{\text{pref}}$ is the value of the Q-tensor preferred by the anchoring at the substrate.

Within $\mathcal{R}$, the fluid density $\rho$, velocity $u$, and Q-tensor evolve over time $t$ according to the continuity, Navier-Stokes, and Beris-Edwards~\citep{Beris94} equations.

\begin{equation}
\partial_{t}\rho+\partial_{\beta}(\rho u_{\beta})=0      
\label{eqn:continuity}
\end{equation}
\begin{multline}
\rho\left(\partial_{t}+u_{\beta}\partial_{\beta}\right)u_{\alpha}=\\
\partial_{\beta}\left[2\mu\Lambda_{\alpha\beta}-p\delta_{\alpha\beta}+\left\{\zeta\Sigma_{\alpha\beta\gamma\delta}+ 
 \mathrm{T}_{\alpha\beta\gamma\delta}\right\}H_{\gamma\delta}\right]\\
 -H_{\beta\gamma}\partial_{\alpha}Q_{\gamma\beta}
\label{eqn:navierStokes} 
\end{multline}
\begin{equation}
\left(\partial_{t}+u_{\gamma}\partial_{\gamma}\right)Q_{\alpha\beta}=
-\zeta\Sigma_{\alpha\beta\gamma\delta}\Lambda_{\gamma\delta}-\mathrm{T}_{\alpha\beta\gamma\delta}\Omega_{\gamma\delta}+\Gamma H_{\alpha\beta} 
\label{eqn:berisEdwards}
\end{equation}

with

\begin{equation}
\begin{split}
H_{\alpha\beta}&=\frac{1}{3}\frac{\delta \mathcal{F}}{\delta Q_{\gamma\gamma}}\delta_{\alpha\beta}-\frac{1}{2}\left(\frac{\delta \mathcal{F}}{\delta Q_{\alpha\beta}}+\frac{\delta \mathcal{F}}{\delta Q_{\beta\alpha}}\right),\\
\Sigma_{\alpha\beta\gamma\delta}&=\tfrac{4}{3}S_{\text{nem}}^{-1}Q_{\alpha\beta}Q_{\gamma\delta}-\delta_{\alpha\gamma}(Q_{\delta\beta}+\tfrac{1}{2}S_{\text{nem}}\delta_{\delta\beta})\\
& -(Q_{\alpha\delta}+\tfrac{1}{2}S_{\text{nem}}\delta_{\alpha\delta})\delta_{\gamma\beta}+\tfrac{2}{3}\delta_{\alpha\beta}(Q_{\gamma\delta}+\tfrac{1}{2}S_{\text{nem}}\delta_{\gamma\delta}),\\
\mathrm{T}_{\alpha\beta\gamma\delta}&=Q_{\alpha\gamma}\delta_{\beta\delta}-\delta_{\alpha\gamma}Q_{\beta\delta},\\
\Lambda_{\alpha\beta}&=\tfrac{1}{2}\left(\partial_{\beta}u_{\alpha}+\partial_{\alpha}u_{\beta}\right),\\
\Omega_{\alpha\beta}&=\tfrac{1}{2}\left(\partial_{\beta}u_{\alpha}-\partial_{\alpha}u_{\beta}\right),
\end{split}
\end{equation}

\noindent where $p=\rho/3$ is the isotropic fluid pressure, $\mu$ is the dynamic viscosity, and $\Gamma$ is the mobility of the nematic order. $\zeta$ is a dynamical parameter, 
dependent on the molecular details of the liquid crystal, that determines how the nematic orientation couples to shear. 
If $\zeta<1$, then the director will tumble indefinitely in the shear, while the case $\zeta>1$ (which we shall consider 
in this paper{\footnote{We anticipate that steady solutions may not exist in the $\zeta<1$ case for sufficiently strong pressure gradients.}}) permits a bulk, 
state-state orientation of the director relative to the shear at the so-called Leslie angle~\citep{leslie68}, which is given by 

\begin{equation}
\thalign=\tfrac{1}{2}\mathrm{arcsec}\zeta.
\end{equation}

At the substrate $\mathcal{W}$, non-slip and anchoring conditions apply,
\begin{align}
(\delta_{\alpha\beta}-\nu_{\alpha}\nu_{\beta})u_{\beta}&=0, \label{eqn:nonslip} \\
L\nu_{\gamma}\partial_{\gamma}Q_{\alpha\beta}&=\alpha\left(Q_{\alpha\beta}^{\text{pref}}-Q_{\alpha\beta}\right),  \label{eqn:substrateEquilibrium}
\end{align}
where $\pmb{\nu}$ is the inward normal to the substrate. As an alternative to Eqn.~(\ref{eqn:substrateEquilibrium}), we may impose the boundary 
condition $\mathbf{Q}=\mathbf{Q}^{\text{pref}}$. This is equivalent to setting $\alpha=\infty$.

We simulate the dynamics of the fluid by discretising space and time - the former into a cubic grid of nodes - and maintain $\rho$, $\mathbf{u}$ and $\mathbf{Q}$ as continuous quantities. 
We utilise a hybrid method in which Eqns.~(\ref{eqn:continuity},\ref{eqn:navierStokes}) are iterated using a lattice Boltzmann method and Eqn.~(\ref{eqn:berisEdwards}) by 
a finite-difference method~\citep{marenduzzo07}, a method that has been previously used by our group when considering a nematic liquid crystal in contact with a 
substrate patterned with rectangular grooves that may fill without the occurrence of complete wetting \cite{Blow2013}. In this study this numerical code was used to 
analyse the dynamical response of the system to an externally-applied electric field so as to identify switching transitions between these filled states. 
As further validation of the numerical code, we chose the simple case of 2D shear flow of nematic liquid crystals for which the exact solution for the 
director profile is known (see Appendix). 

We choose the parameters (in simulation units) $A=0.5$, $\tau^{*}=0.9$, $L=0.033(3)$, $\Gamma=0.025$, $\rho=80$ (with deviations due to the effects of compressibility, which are small), 
$\mu=13.33(3)$, $\zeta=1.5$ (aligning regime) and $S_{\text{nem}}=1$. $p$, $\alpha$ and $\mathbf{Q}^{\text{pref}}$ are the parameters of interest that we shall vary.

\begin{figure}
\centering
\includegraphics[width=80mm]{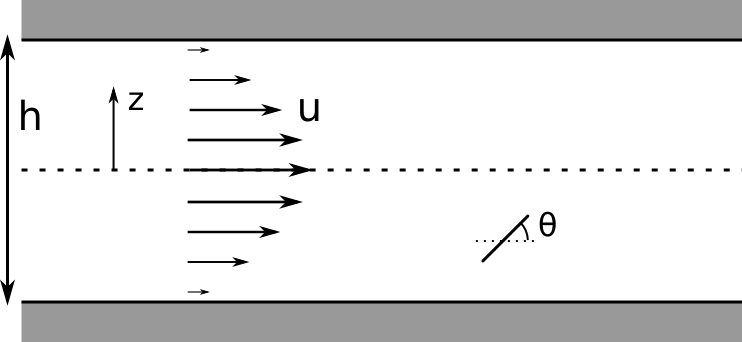}
\caption{Illustration of the channel dimensions, flow profile and director angle.}
\label{fig:channelBasic}
\end{figure}
We consider the system geometry depicted in figure~(\ref{fig:channelBasic}), where the substrate is comprised of two parallel plates that lie in the $xy$ plane at $z=-h/2$ and $z=h/2$. In the $y$ direction, the system has a width of only one node, imposing uniformity in this direction. 
In the $x$ direction, the system extends from $x=0$ to $x=l$, and at these boundaries we impose a Neumann condition $\partial_{x}=0$ on $\mathbf{Q}$ and $\mathbf{u}$, and a pressure 
difference condition $p=p_{0}+\delta p/2$ at $x=0$ and $p=p_{0}-\delta p/2$ at $x=l$, which is achieved by fixing the density at these coordinates according to the relation $p=\rho/3$ 
(the maximum density difference used in our simulations is $\delta \rho/\rho=0.011$, so compression effects are negligable). Thus, we are simulating two-dimensional channel flow, 
in a channel of width $h$, subjected to a pressure gradient $G=\delta p/l$. We use $l=100$ and $h=24$. Since we find that $u_{z}$ remains very small in all our simulations, 
we hereon denote $u_{x}$ by the indexless $u$.

In order to simplify our description of the system and comparison with theory, we define the following rescaled quantities
\begin{eqnarray}
\tz&=\frac{2z}{h}&\text{dimensionless length} \nonumber\\
\tG&=\frac{G h^{3}}{4L}&\text{dimensionless pressure gradient} \nonumber\\
\ta&=\frac{\alpha h}{2L}&\text{dimensionless anchoring}\nonumber\\
\tm&=\frac{\mu h}{L}&\text{isotropic viscosity rescaled to} \nonumber\\
& &\text{units of inverse velocity} \nonumber\\
r&=\frac{1}{\mu\Gamma};&\text{degree of anisotropy in the viscosity}
\end{eqnarray}

We note that the quantity $\tm r U$, where $U$ is a characteristic flow velocity, corresponds to what is commonly defined as the Ericksen number~\cite{erickson69}.

\section{Results}
\label{sec:results}

\subsection{Homeotropic anchoring}
\label{sec:homeotropic}

\begin{figure*}
\centering
\includegraphics[width=120mm]{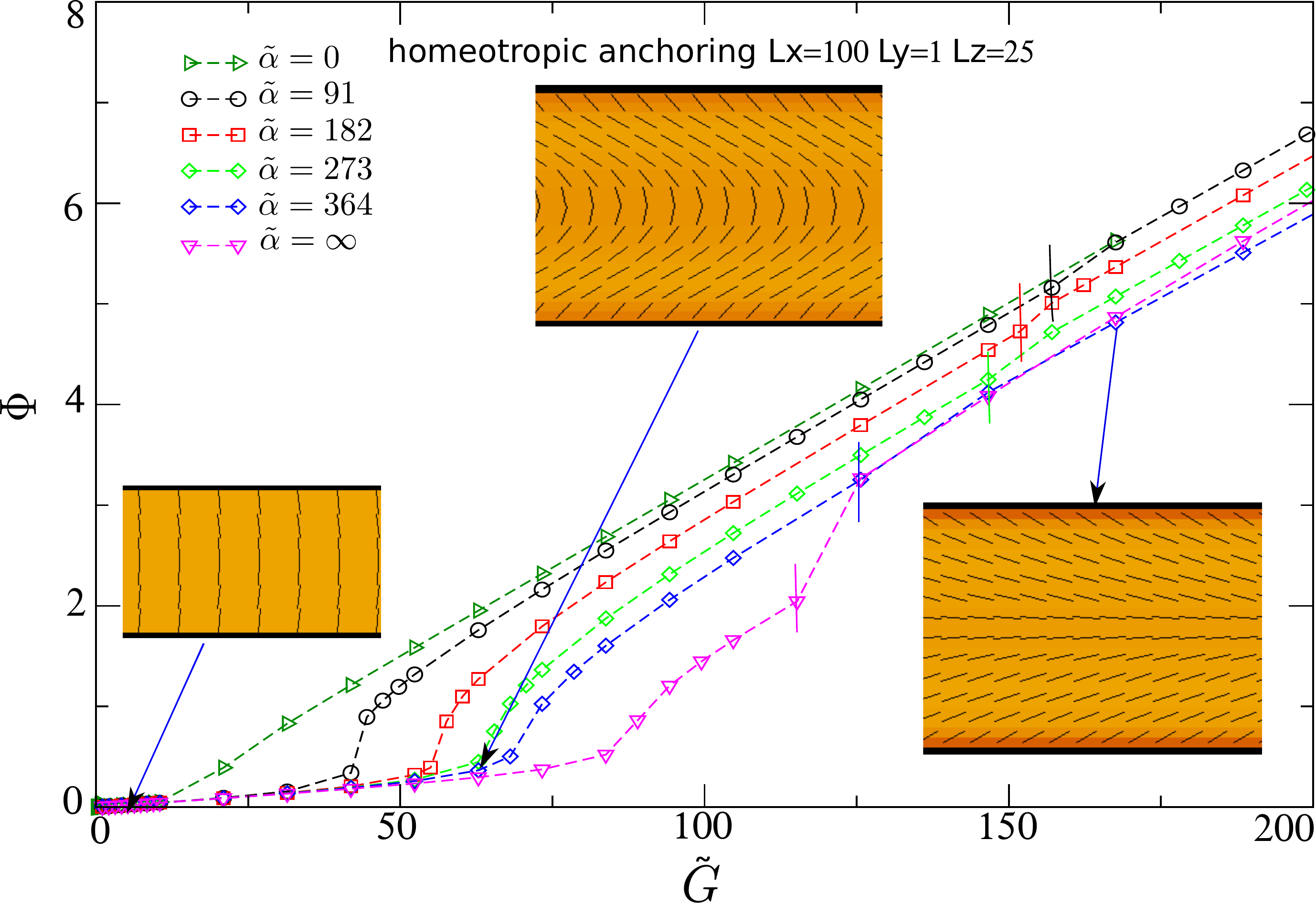}
\caption{Mass flow rate $\Phi$ as a function of the pressure gradient via simulation. Different curves correspond to 
distinct values for the homeotropic anchoring condition. For $\ta=364$, an example case, the director orientation (black lines) obtained 
for different pressures gradients are shown in the embedded figures for the case of the 'vertical' state [at low $\tG$ (lower left) and near the 
jump in flow rate (top centre)] and 'horizontal' state (lower right). Vertical lines represent the value of $\tG$ above which we observe the 'horizontal' state.}
\label{fig:FluxAlpha}
\end{figure*}

We first consider the case where both channel walls have homeotropic (i.e. director perpendicular to the wall) anchoring conditions. Obviously, in 
the absence of flow, the director will uniformly orient itself in the $z$ direction. Following the terminology of \citet{jewell09}, we call this configuration 
the `vertical' state, $\mathbb{V}$. 

For various values $\tG$ and $\ta$, we evolve the system until a steady state is obtained. Depending on the driving $\tG$, simulations were 
run up to $1\times10^6$ time-steps. However, in our experience, systems driven at low $\tG$ may take $3$ to $4$ times longer than high $\tG$ 
to reach a steady state.

Once steady state is achieved, we measure the mass flow rate, defined as
\begin{equation}
\Phi=\int^{h/2}_{-h/2}\rho u \dd z =\frac{1}{2}\int^{1}_{-1}\rho h u \dd\tz. \label{eqn:flux}
\end{equation}

In figure~(\ref{fig:FluxAlpha}) we plot $\Phi$ as a function of $\tG$ for various anchoring strengths ranging from $\ta=0$ to $\infty$. Symbols represent simulation 
data and discontinuous lines are added to help guide the eye. The insets show the nematic configuration recorded for the indicated data points 
for the case $\ta=364$, a relatively strong anchoring. The orange shading represents the nematic order parameter $S$ (which, in all cases, varies only 
slightly away from $S_{\text{nem}}$), and the black lines depict the director $\mathbf{n}$.

Examining figure~(\ref{fig:FluxAlpha}), we see that for small $\tG$, the director is perturbed only slightly from the uniform vertical configuration, 
as shown by the leftmost inset in figure~(\ref{fig:FluxAlpha}). In this region of the graph, the increase of $\Phi$ with $\tG$ is gradual. However, as $\tG$ increases, 
there comes a point where $\Phi$ undergoes a rapid increase with respect to $\tG$. The value of $\tG$ at which this departure from linearity occurs is higher for systems 
with larger $\ta$. Shortly after this jump, $d\Phi/d\tG$ moderates, but remains at a higher value than before the jump. As $\tG$ is increased further, the relation is 
approximately linear, albeit with a non-zero intercept. For the case of $\ta=0$, instead of a jump there is a kink, in which $d\Phi/d\tG$ switches from a lower to a higher value.

The middle inset illustrates the nematic texture prior to the jump. We see that the nematic texture is strongly distorted, with the director 
significantly perturbed from the vertical away from the walls and channel centre. However, the state remains topologically equivalent to that depicted in the left inset, 
since one state can be continuously deformed into the other. Therefore we also class the state in the middle inset as $\mathbb{V}$.

As $\tG$ is increased further, the director profile undergoes an abrupt transition so that the director orientation at the middle of the channel becomes horizontal. 
Again following \citet{jewell09} we term this new state the horizontal state, $\mathbb{H}$. The change $\mathbb{V}\rightarrow\mathbb{H}$ represents a topological change 
and cannot be effected by continuous deformations. As noted by \citet{denniston01CTPS}, the transition occurs by the formation of topological defects, 
which unwind to produce the new state. Intriguingly, the transition only causes a small discontinuity in the dependence of $\Phi$ on $\tG$, barely noticeable in 
comparison to the jump described in the previous two paragraphs. Contrary to what was seen with the jump, we observe that the $\mathbb{V}\rightarrow\mathbb{H}$ transition occurs 
at smaller values of $\tG$ for larger $\ta$, as can be observed by the vertical discontinuous lines in figure~(\ref{fig:FluxAlpha}).

In order to check the simulation results and to gain understanding of their key features, we now perform calculations based on the fluid equations. Since the degree of 
nematic order remains roughly constant at $S_{\text{nem}}=1$, and since the director is confined to the $zx$ plane, we may write
\begin{align}
Q_{xx}&=\tfrac{1}{4}\left(3\cos 2\theta+1\right),\nonumber\\
Q_{zz}&=\tfrac{1}{4}\left(-3\cos 2\theta+1\right),\\
Q_{zx}&=\tfrac{3}{4}\sin 2\theta,\nonumber
\end{align}
$\theta$ being the anticlockwise angle made by the director with the $x$ axis. Furthermore, the simulations verify that the $z$ component of velocity is negligible, 
the fluid is approximately incompressible, and $u$ and $\theta$ show no variation in the $x$ direction. Under these assumptions, Eqn.~(\ref{eqn:continuity}) becomes 
trivial, and Eqns.~(\ref{eqn:navierStokes},\ref{eqn:berisEdwards},\ref{eqn:substrateEquilibrium}) may be reduced to
\begin{align}
&\frac{\dd}{\dd\tz}\left(\tm\frac{\dd u}{\dd\tz}-4\left(\zeta\cos2\theta-1\right)\frac{\dd^{2}\theta}{\dd\tz^{2}}\right)=-\tG,   \label{eqn:navierStokesDimless} \\
&\frac{\dd^{2}\theta}{\dd\tz^{2}}= -\frac{1}{4}r\tm\left(\zeta\cos 2\theta -1\right)\frac{\dd u}{\dd\tz}, \label{eqn:berisEdwardsDimless} \\
&\frac{\dd\theta}{\dd\tz}=\pm \tfrac{1}{2}\ta\sin2\theta\;\text{at}\;\tz=\pm 1.  \label{eqn:subtrateEquilibriumDimless}
\end{align}

We integrate Eqn.~(\ref{eqn:navierStokesDimless}) once, and use the condition of symmetry that $\dd u/\dd\tz=0$ at the channel midpoint $\tz=0$. We then 
rearrange Eqns.~(\ref{eqn:navierStokesDimless},\ref{eqn:berisEdwardsDimless}) to separate the derivatives of the two variables, giving
\begin{align}
\frac{\dd u}{\dd\tz}&=-\frac{\tG}{\tm\left(1+r\left(\zeta\cos 2\theta-1\right)^{2}\right)}\tz,   \label{eqn:velocityODE} \\
\frac{\dd^{2}\theta}{\dd\tz^{2}}&= \frac{r\tG\left(\zeta\cos 2\theta -1\right)}{4\left(1+r\left(\zeta\cos2\theta-1\right)^{2}\right)}\tz. \label{eqn:angleODE} 
\end{align}

Examining Eqn.~(\ref{eqn:velocityODE}), we note that the velocity gradient becomes larger as $\theta$ approaches $\thalign$. 
Thus, as the director orientation becomes increasingly distorted away from the vertical orientation by the flow, we expect a boost to $u$, and 
hence to $\Phi$, compared to a hypothetical situation where the director orientation remains vertical. $d\Phi/d\tG$ therefore has a regular contribution 
as would be found in Poiseuille's law, and an additional contribution due to the changes in the profile of $\theta$. The jump 
observed in figure~(\ref{fig:FluxAlpha}) corresponds to especially rapid changes in $\theta$.

We now seek to confirm the results for small $\tG$, by finding a series solution for Eqns.~(\ref{eqn:velocityODE},\ref{eqn:angleODE}). In such a limit, $u$ will be small, 
and provided that the anchoring is not too weak, the nematic orientation will be perturbed only mildly from the vertical (as shown in the red curve of 
figure~\ref{fig:switchTopology}). We may therefore write expansions for $\theta$ and $u$.
\begin{align}
u&=u_{1}\tG + u_{3}\tG^{3} +...\\
\theta&=\tfrac{\pi}{2}+\theta_{1}\tG+\theta_{3}\tG^{3}+...
\label{eqn:LowFlowTheta}
\end{align}
We include only odd powers because symmetry dictates that $u \rightarrow -u$ and $\theta-\pi/2\rightarrow-(\theta-\pi/2)$ under $\tG \rightarrow -\tG$

In the lowest order, Eqns.~(\ref{eqn:velocityODE},\ref{eqn:angleODE},\ref{eqn:nonslip},\ref{eqn:subtrateEquilibriumDimless}) become
\begin{align}
\tm\left(1+r\left(\zeta+1\right)^{2}\right)\frac{\dd u_{1}}{\dd\tz}&=-\tz,  \\
\frac{\dd^{2}\theta_{1}}{\dd\tz^{2}}&= -\frac{r\left(\zeta+1\right)}{4\left(1+r\left(\zeta+1\right)^{2}\right)}\tz, \\
u_{1}&=0\;\text{at}\;\tz=\pm 1,\\ 
\frac{\dd\theta_{1}}{\dd\tz}&=\mp\ta\theta_{1}\;\text{at}\;\tz=\pm 1,
\end{align}
the solutions of which are
\begin{align}
u_{1}&=\frac{1}{2\tm\left(1+r\left(\zeta+1\right)^{2}\right)}\left(1-\tz^{2}\right),  \\
\theta_{1}&=\frac{r\left(\zeta+1\right)}{24\left(1+r\left(\zeta+1\right)^{2}\right)}\left\{\left(\frac{\ta+3}{\ta+1}\right)\tz-\tz^{3}\right\}.
\end{align}

\begin{figure}
\centering
\includegraphics[width=70mm]{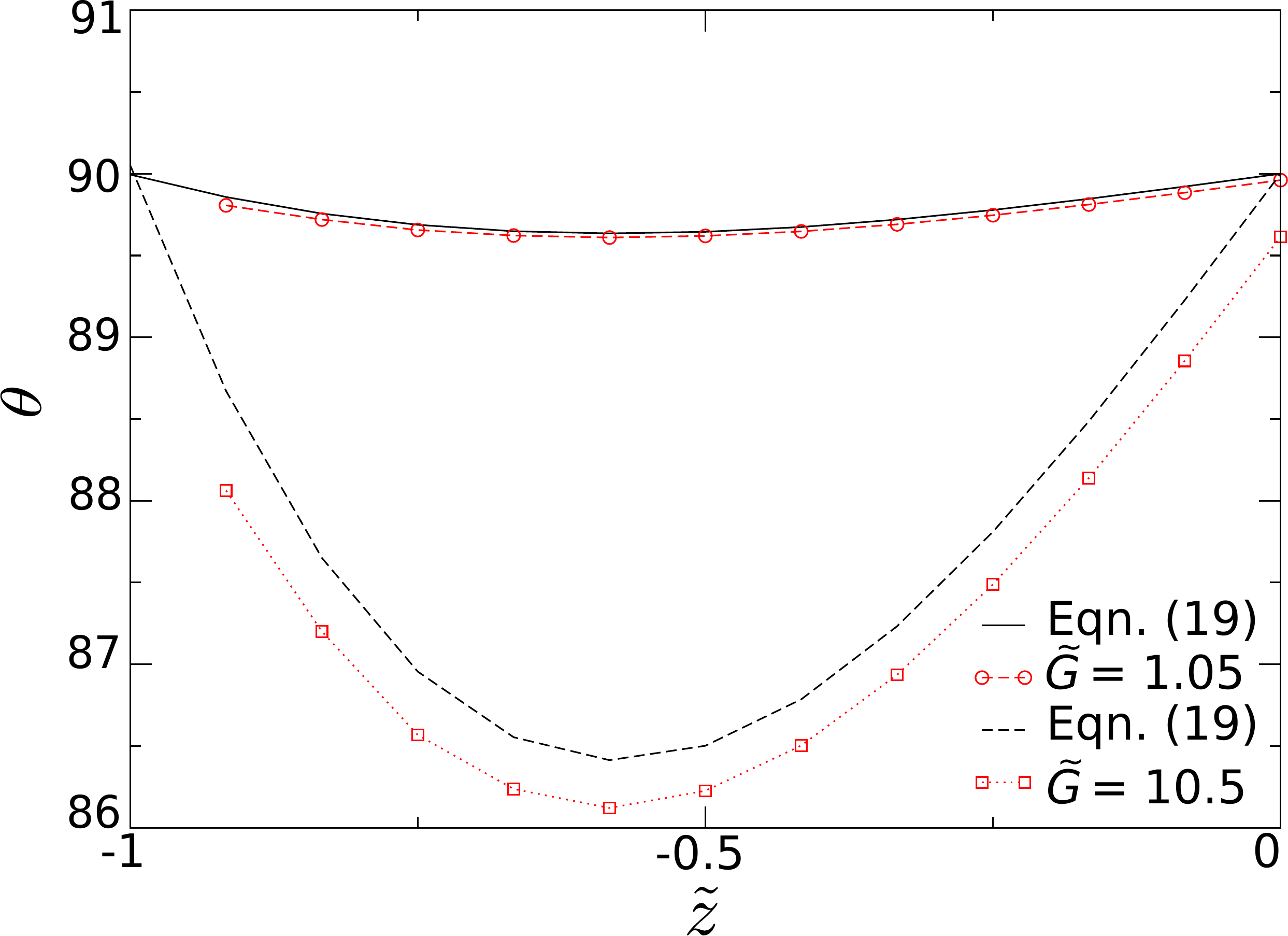}
\caption{Director angle $\theta$ across the half of the width of the channel. Black lines are angles obtained using 
Eqn.~(\ref{eqn:LowFlowTheta}) whilst simulation results are represented by red symbols connected by a red dashed line to help guide the eye. 
Inset: Illustrates the deviation of numerical results from theory as one expects when increasing $\tG$.}
\label{fig:lowFlowThetaSIM}
\end{figure}

The $\tz$ dependence of $u_{1}$ and $\theta_{1}$ is in agreement with \citet{zhou07} (where only the strong anchoring limit is considered). $u_{1}$ is 
the standard Poiseuille form, but with a modified coefficient. It is dependent on the dynamical properties of the liquid crystal, but not on the anchoring. 
By contrast, the director profile does have dependence on $\ta$. It is worth mentioning that we find that the director 
component in the y-direction remains very small in all our simulations. 
Therefore our director is confined to the xz-plane. 
We denote the direction by the angle of the director with respect to the z-axis which we named theta. 
Figure~(\ref{fig:lowFlowThetaSIM}) compares the angle of the director $\theta$ as given 
by Eqn.~(\ref{eqn:LowFlowTheta}) for $\tG=1.05$ and for $\tG$ an order of magnitude larger. The director angle given by Eq.~(\ref{eqn:LowFlowTheta}) 
is represented by black lines. The respective angle of the director from the simulation with $\ta=364$ is shown by red symbols.

As expected, for the lower driving pressure gradient, $\tG=1.05$, we observe only a small deviation of $\theta$ from $\pi/2$, which is in 
very good agreement with that given by Eqn.~(\ref{eqn:LowFlowTheta}). When applying a pressure gradient an order of magnitude larger, 
Eqn.~(\ref{eqn:LowFlowTheta}) overestimates $\theta$ but is still in good agreement with a relative error $< 10\%$.

The next order velocity equation is

\begin{equation}
\begin{split}
\frac{\dd u_{3}}{\dd z}&=-\frac{4r\zeta(\zeta+1)}{\tm\left(1+r\left(\zeta+1\right)^{2}\right)^{2}}\theta_{1}^{2}\tz,   \\
&=-\frac{r^{3}\zeta(\zeta+1)^{3}}{144\tm\left(1+r\left(\zeta+1\right)^{2}\right)^{4}}\\
&\left\{\left(\frac{\ta+3}{\ta+1}\right)^{2}\tz^{3}-2\left(\frac{\ta+3}{\ta+1}\right)\tz^{5}+\tz^{7}\right\}
\end{split}
\end{equation}
giving
\begin{align*}
&u_{3}=\frac{r^{3}\zeta(\zeta+1)^{3}}{144\mu\left(1+r\left(\zeta+1\right)^{2}\right)^{4}}\\
&\left\{\left(\frac{\ta+3}{\ta+1}\right)^{2}\left(\frac{1-\tz^{4}}{4}\right)-\left(\frac{\ta+3}{\ta+1}\right)\left(\frac{1-\tz^{6}}{3}\right)+\frac{1-\tz^{8}}{8}\right\}
\end{align*}

\noindent $\theta_{3}$ may be similarly found, and then used to find the next order contributions, and so on.

The mass flow rate, in the lowest two orders of $\tG$, is given by

\begin{equation}
\begin{split}
\Phi&=\tfrac{1}{2}\rho h \int^{1}_{-1} u d\tz,  \\
&=\rho h \frac{1}{3\tm\left(1+r\left(\zeta+1\right)^{2}\right)}\tG+\rho h\frac{r^{3}\zeta(\zeta+1)^{3}}{144\tm\left(1+r\left(\zeta+1\right)^{2}\right)^{4}} \\
&\left\{\frac{1}{5}\left(\frac{\ta+3}{\ta+1}\right)^{2}-\frac{2}{7}\left(\frac{\ta+3}{\ta+1}\right)+\frac{1}{9}\right\}\tG^{3}+\mathrm{O}(\tG^{5}).
\end{split}
\label{LowFlowFlux}
\end{equation}
We see that the linear term is essentially that of the Poiseuille relation, but with the viscosity modified by the 
factor $\left(1+r\left(\zeta+1\right)^{2}\right)$. This linear term is not dependent on the anchoring strength. The cubic term represents the 
leading order boost to the flow rate as the director begins to distort closer towards $\thalign$. This boost is larger for smaller $\ta$, as would 
be expected since the director profile will distort more easily if the anchoring is weaker.

\begin{figure}
\centering
\includegraphics[width=80mm]{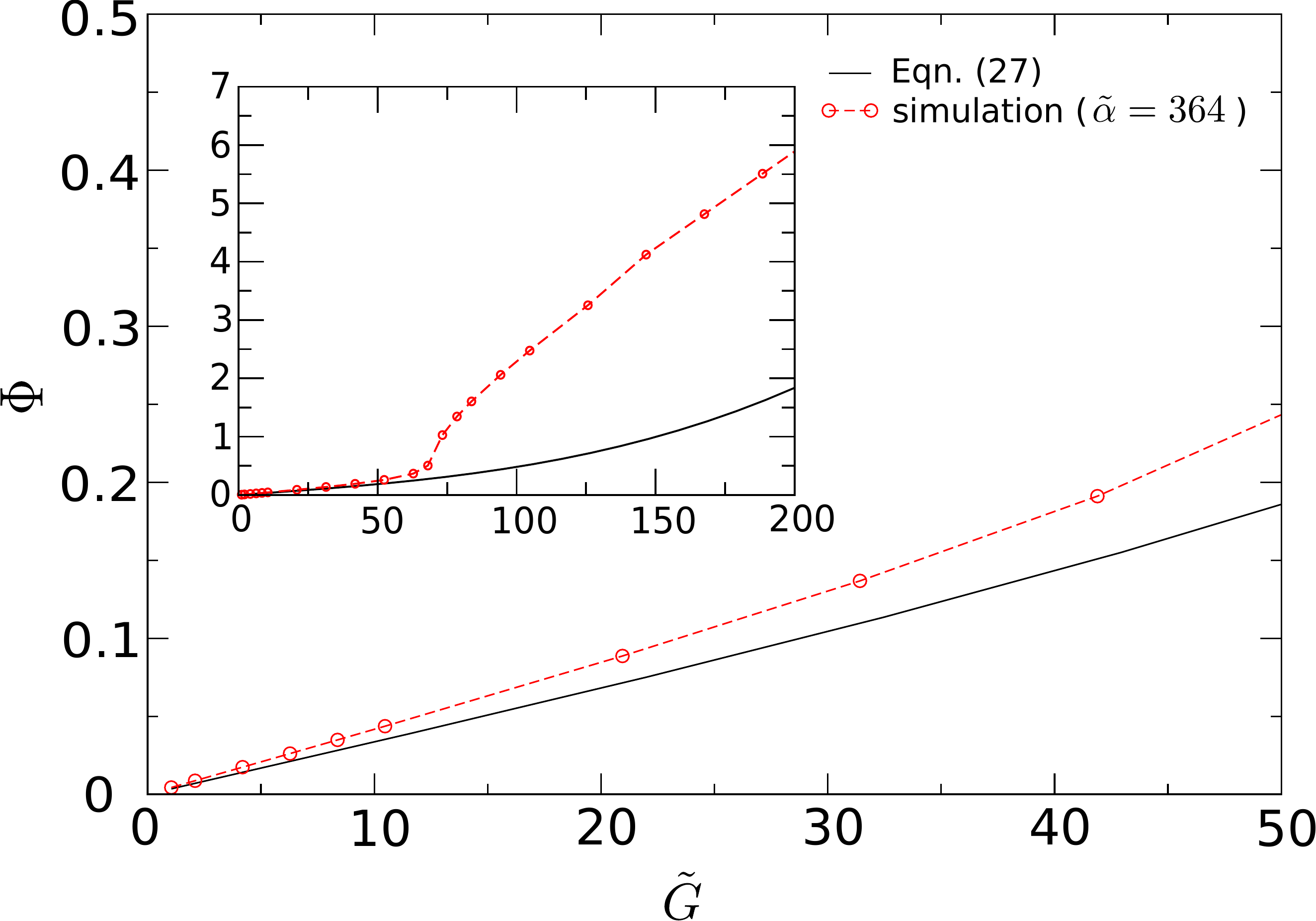}
\caption{Comparison of the mass flow rate $\Phi$ as a function of pressure gradient obtained using Eqn.~(\ref{LowFlowFlux}) and the corresponding simulation results. 
Embedded graph illustrates the expected deviation of the numerical results from theory as $\tG$ increases.}
\label{fig:lowFlowFluxSIM}
\end{figure}

\begin{figure*}
\centering
\includegraphics[width=120mm]{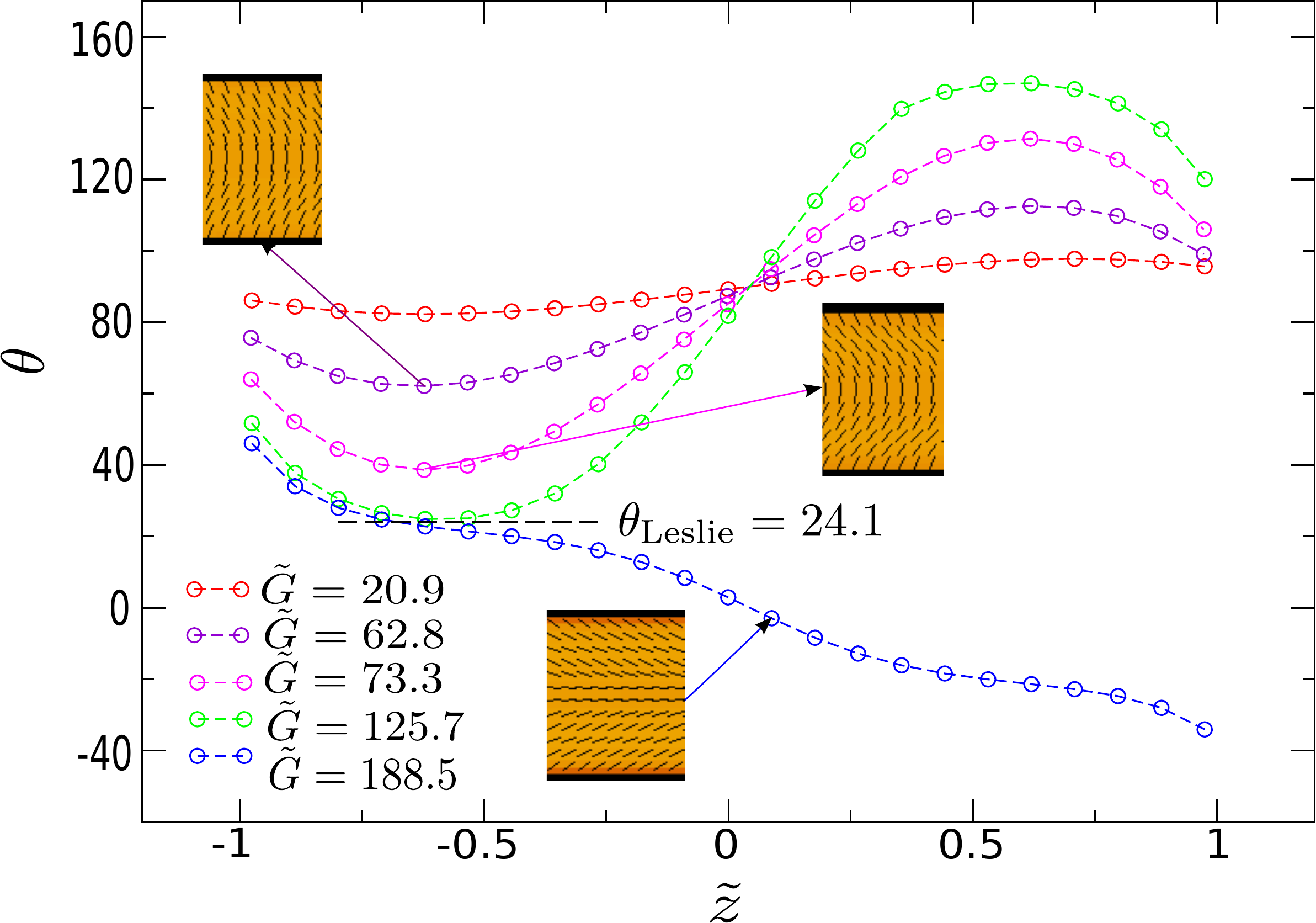}
\caption{Graph illustrating the switch from the 'vertical' to 'horizontal' texture. In this figure the director angle $\theta$ across the width 
of the channel when $\tG$ is small (red) and the crossover from the 'vertical' to 'horizontal' when $\theta$ approaches the alignment angle (green to blue). 
As examples, the director orientation (black lines) obtained for different pressures gradients are shown in the embedded figures.}
\label{fig:switchTopology}
\end{figure*}

As an example, figure~(\ref{fig:lowFlowFluxSIM}) compares the simulation data (connected red points) against the prediction of 
Eqn.~(\ref{LowFlowFlux}), $\ta=364$. For low $\tG$, the mass flow rate of the channel is very well represented by Eqn.~(\ref{LowFlowFlux}). 
This agreement is seen close up to values of $\tG$ before the onset of the sudden jump in $\Phi$, as illustrated in the inset of this figure.

\begin{figure}
\centering
\includegraphics[width=80mm]{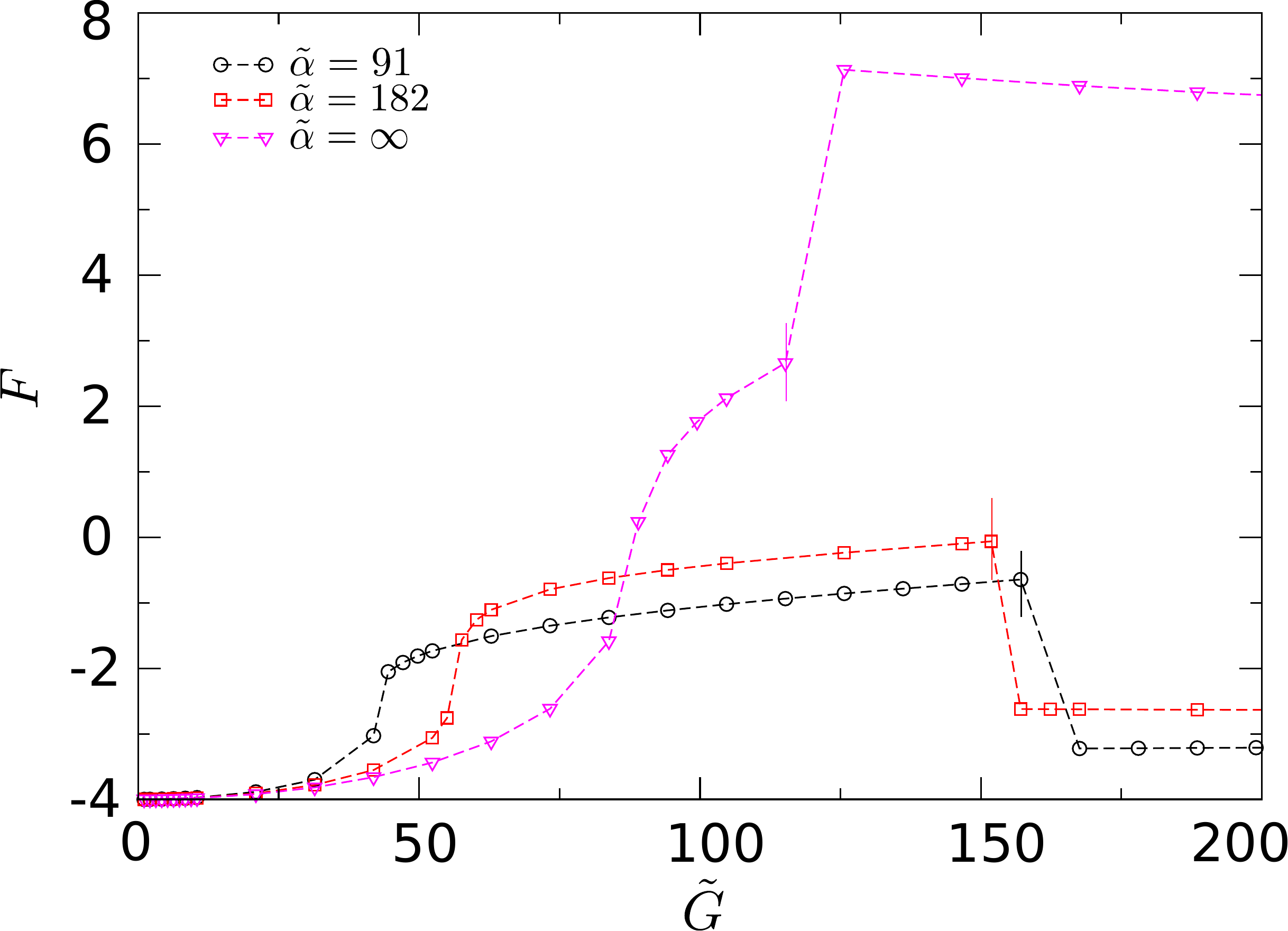}
\caption{Free energy as a function of the pressure gradient via simulation. 
Different curves correspond to distinct values of $\ta$ for the homeotropic anchoring condition. Vertical lines represent the 
value of $\tG$ above which we observe the 'horizontal' state.}
\label{fig:FreeEAlpha}
\end{figure}

Next, we elucidate the cause of the $\mathbb{V}\rightarrow\mathbb{H}$ transition. In $\mathbb{V}$, illustrated by the red and green curves 
in figure~(\ref{fig:switchTopology}), the director angle $\theta$ starts at $\thsub$ (the actual angle given by Eqn.~(\ref{eqn:subtrateEquilibriumDimless}), 
not the angle $\pi/2$ ideally preferred by the anchoring) at the substrate $\tz=-1$, decreases down to some value $\thturn$, then 
increases to $\pi/2$ at the channel midpoint. In the upper portion of the channel, the profile is inverted, with $\theta$ reaching a 
maximum of $\pi-\thturn$, and arriving at the upper substrate $\tz=1$ with $\pi-\thsub$. Thus the integrated change in angle across the 
channel is $\Delta\theta=\pi-2\thsub$. By contrast, in $\mathbb{H}$, depicted by the blue curve in figure~(\ref{fig:switchTopology}), 
$\theta$ decreases monotonically across the channel, starting at $\thsub$ (generally with a different value to the previous case) at $\tz=-1$, 
decreasing through $\theta=0$ at $\tz=0$, and arriving at $\tz=1$ with the angle $-\thsub$, thus achieving an integrated change 
of $\Delta\theta=-2\thsub$. The instability by which the vertical state gives way to the horizontal has a dynamical cause, as we shall now show.

Examining Eqn.~(\ref{eqn:angleODE}), we note that, in the lower (upper) segment $\tz<0$ ($\tz>0$), $d^{2}\theta/d\tz^{2}$ is positive (negative) 
when $\cos2\theta<\cos2\thalign$, and vice-versa when $\cos2\theta>\cos2\thalign$. Thus, in the case where $\thturn>\thalign$, $d^{2}\theta/d\tz^{2}$ is 
positive throughout the bottom segment (and negative in the upper segment), including at $\thturn$, which is therefore a local minimum in $\theta$. However, 
if $\theta$ falls below $\thalign$, $d^{2}\theta/d\tz^{2}$ becomes negative. It now becomes impossible for $\theta$ to reach a turning point and increase to 
meet $\pi/2$ at 0. Thus, the transition from the vertical to horizontal state occurs when $\thturn$ exceeds $\thalign$. For the value $\zeta=1.5$ in our system, 
we derive $\thalign=24.1^{\circ}$. This value is in good agreement with the transition angle found by our simulations, as shown in figure~(\ref{fig:switchTopology}).

Our findings indicate that the switch is driven by dynamical considerations, under which $\mathbb{V}$ becomes unstable in the flow. Figure~(\ref{fig:FreeEAlpha}) shows 
that the free energy, Eqn.~\ref{eqn:freeEnergy}, typically changes discontinuously at the transition, with a decrease in the free energy observed for all measured 
values of $\ta$ except $\ta=\infty$. This contradicts the explanation of~\citet{jewell09} that the transition occurs at the point where the free energies of the two 
textures are equal, and supports the observation of~\citet{denniston01CTPS} that $\mathbb{V}$ can persist as a metastable state; we find that it does so until driven 
into the $\mathbb{H}$ state by the dynamical instability. The $\ta=\infty$ case shows an increase in free energy, indicating that the $\mathbb{V}$ state becomes 
dynamically unstable before any free energy crossover with $\mathbb{H}$ occurs.

\begin{figure}
\centering
\includegraphics[width=80mm]{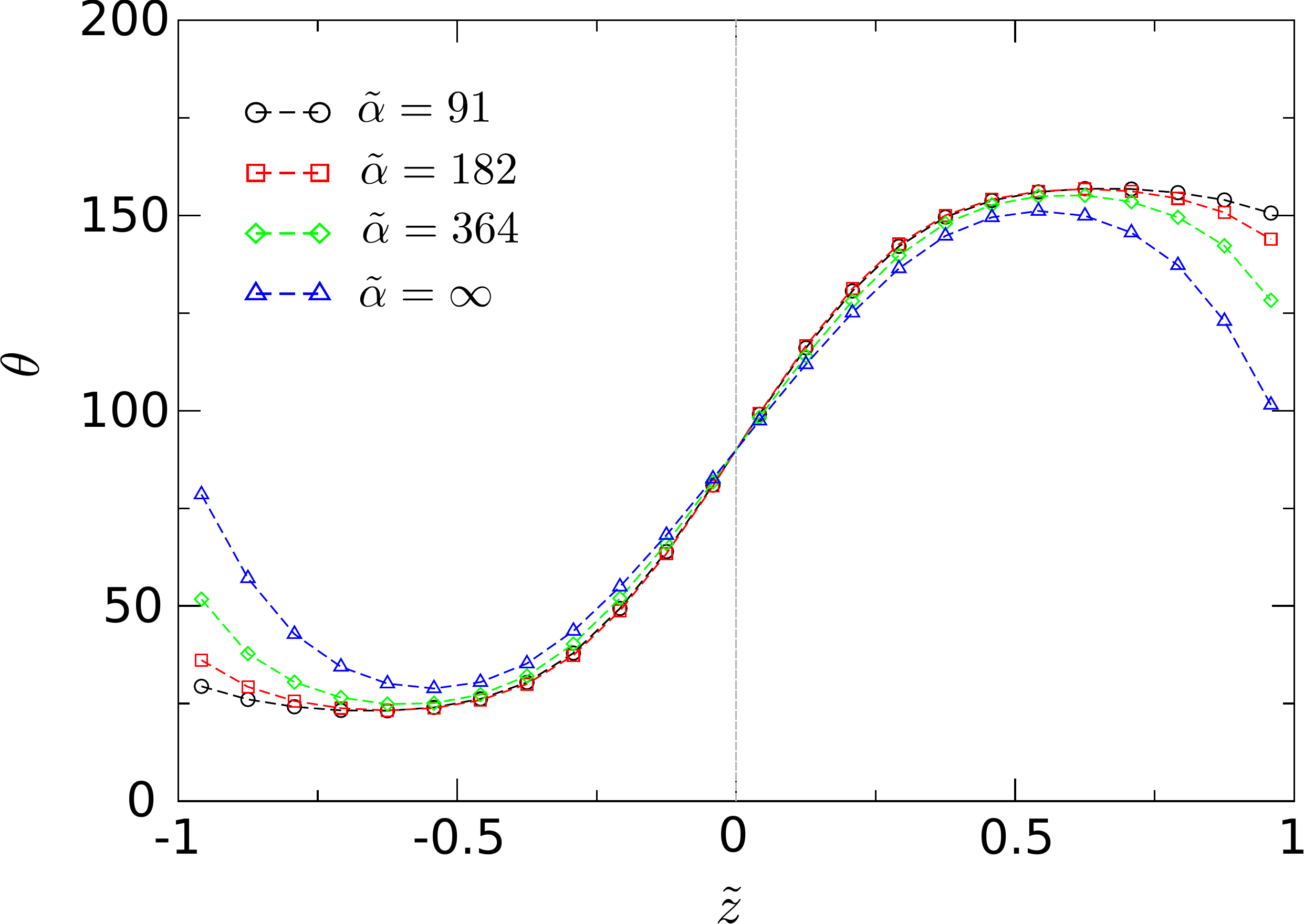}
\caption{Director profile as a function of channel height $\tz$ near the $\mathbb{V}\rightarrow\mathbb{H}$ transition. 
Different curves correspond to distinct values of $\ta$ for the homeotropic anchoring condition. Vertical discontinuous grey line indicated channel at mid-height.}
\label{fig:ThetaVH}
\end{figure}

In light of this explanation, the fact that the $\mathbb{V}\rightarrow\mathbb{H}$ transition requires larger $\tG$ for {\em smaller} $\ta$ may seem counter-intuitive. 
If a weaker anchoring provides less resistance to distortions of the director away from the vertical, then shouldn't less forcing be required to make $\thturn$ fall 
below $\thalign$? The answer to the paradox lies in the substrate boundary condition, Eqn.~(\ref{eqn:subtrateEquilibriumDimless}), which shows that $\ta$ is an amplifying factor for 
the magnitude of $\dd\theta/\dd\tz$ at the substrate. When $\ta$ is large, this derivative is large at the substrate (and negative, assuming that we are considering the 
lower segment of the channel). Conversely, in the limit of $\ta\rightarrow 0$, the substrate derivative vanishes. Since $d^{2}\theta/d\tz^{2}$ is positive for $\theta>\thturn$, 
we see that in this weak anchoring limit $\dd\theta/\dd\tz$ immediately becomes positive away from the substrate, and therefore $\theta$ cannot fall below $\thalign$ unless $\thsub<\thalign$ 
to begin with, which requires a large value of $\tG$ to achieve. We thus see that weak anchoring, by limiting the initial derivative of $\theta$ at the substrate 
(as can be seen in figure (\ref{fig:ThetaVH})), impedes rather than assists the $\mathbb{V}\rightarrow\mathbb{H}$ transition.

Another counter-intuitive aspect of our findings is that the jump shown in figure~(\ref{fig:FluxAlpha}), which occurs within the state $\mathbb{V}$, should constitute a much 
larger departure from non-linearity than the transition $\mathbb{V}\rightarrow\mathbb{H}$, which involves a discontinuous change of the nematic texture. To resolve this puzzle, 
we again examine Eqn.~(\ref{eqn:velocityODE}), which shows how $\theta$ influences the gradient $du/d\tz$. The corresponding flow rate $\Phi$ is found by 
performing two successive integrations, from the walls where $u$ is fixed at zero to the centre. Thus changes in $\theta$ close to the walls will have a greater 
effect on $\Phi$ than changes in $\theta$ close to the centre of the channel. We see from figure~(\ref{fig:switchTopology}), that in the 
transition $\mathbb{V}\rightarrow\mathbb{H}$ (from the green to blue curves), there is a little change in the director profile near the walls, and change is 
limited towards the centre of the channel. By contrast, during the jump in $\mathbb{V}$ (purple to pink curves), there is significant change in $\theta$ 
across the entire width of the channel.

\begin{figure}
\centering
\includegraphics[width=80mm]{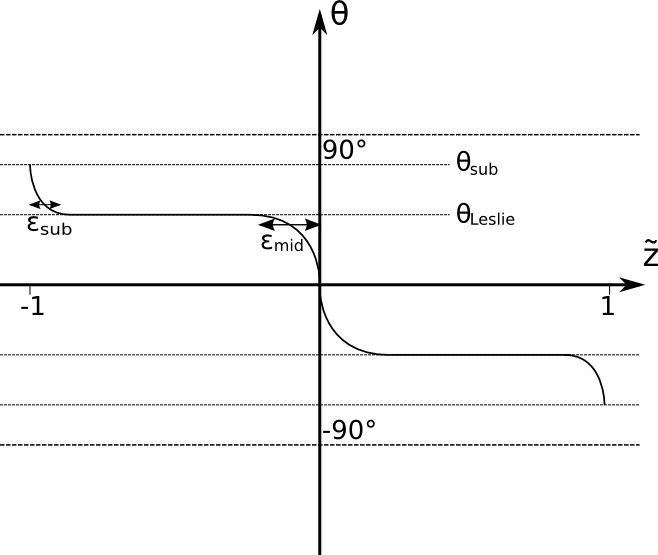}
\caption{Schematic graph of the director angle $\theta$ across the width of the channel when the pressure gradient is large.}
\label{fig:hiFlow}
\end{figure}

\begin{figure}
\centering
\includegraphics[width=80mm]{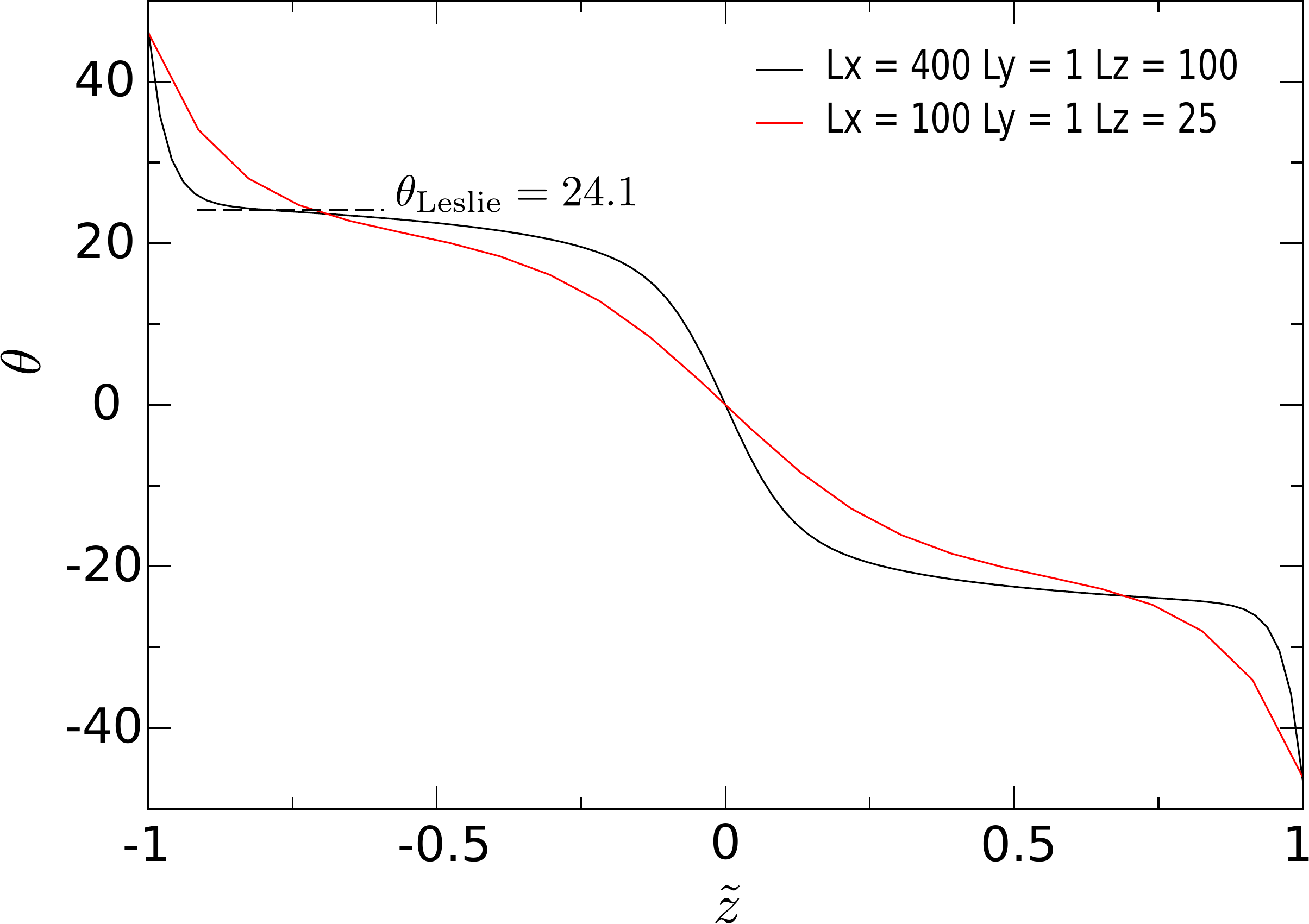}
\caption{Director angle $\theta$ across the width of the channel with homeotropic anchoring of strength $\ta=1500$ obtained two systems with $l=400,h=99$ and $l=100,h=24$ ).}
\label{fig:HighFluxTheta}
\end{figure}

Following the work of~\citet{leslie66} and~\citet{degennes93}, in the limit of high $\tG$ we except $\theta$ to be close to $\pm\thalign$ for 
most of the channel width, deviating only in layers at the edges and middle of the channel, as shown in figure~(\ref{fig:hiFlow})

\begin{equation}
\theta=\begin{cases}
\text{varies rapidly from $[\thsub,\thalign]$}:-1<\tz<-1+\epsub\\
\text{steady at $\thalign$}:\;\;\;\;\;\;\;\;\;\;\;\;\;\;\;\;\;\;\;\;-1+\epsub<\tz<-\epmid\\
\text{varies rapidly from}\\
\text{[$-\thalign,0[$ through $[0,\thalign]$}:\;\;\;\;\;\;\;\;-\epmid <\tz<\epmid\\
\text{steady at $-\thalign$}:\;\;\;\;\;\;\;\;\;\;\;\;\;\;\;\;\;\;\;\;\;\;\;\;\;\;\;\;\;\;\;\;\;0 < \tz <\epmid\\
\text{varies rapidly from $[-\thalign,-\thsub]$}:1-\epsub<\tz<1.
\end{cases}
\end{equation}

\noindent where $\epmid$ and $\epsub$ are the characteristic thicknesses of the layers at the substrates and middle of the channel respectively. 
Simulations in figure~(\ref{fig:HighFluxTheta}) confirms this prediction.

To calculate an estimate of the flow rate, we solve Eqn.~(\ref{eqn:velocityODE}) for the simplified stepwise function
\begin{equation}
\theta\approx\begin{cases}
\thsub & -1 < \tz <-1+\epsub,  \\
\thalign & -1+\epsub < \tz <-\epmid, \\
0  & -\epmid < \tz <\epmid, \\
-\thalign & 0 < \tz <\epmid, \\
-\thsub & 1-\epsub < \tz <1,
\end{cases}
\end{equation}
where $\thsub$ is measured and $\epsub$ and $\epmid$ estimated from the simulations. We find that

\begin{equation}
u\approx\begin{cases}
\frac{\tG}{2\tm}f(\thsub)(1-\tz^{2})\;\;\;\;\;\;\;\;\;\;\;\;\;\;\;\;\;-1 < \tz <-1+\epsub\\
\\
\frac{\tG}{2\tm}f(\thsub)(1-[1-\epsub]^{2}) \\
+ \frac{\tG}{2\tm}f(\thalign)\\
([1-\epsub]^{2}-\tz^{2}) \;\;\;\;\;\;\;\;\;\;\;\;\;\;\;-1+\epsub < \tz <-\epmid\\
\\
\frac{\tG}{2\tm}f(\thsub)(1-[1-\epsub]^{2})  \\
+ \frac{\tG}{2\tm}f(\thalign)([1-\epsub]^{2}-\epmid^{2})  \\
+ \frac{\tG}{2\tm}f(0)(\epmid^{2}-\tz^{2})\;\;\;\;\;\;\;\;\;\;\;\;\;\;\;\;\;\;\;\;\;\;\;-\epmid < \tz < 0
\end{cases}
\end{equation}

\noindent and, as the velocity profile is mirrored in the upper half of the channel, where
\begin{equation}
f(\theta)=\frac{1}{1+r(\zeta\cos 2\theta -1)^{2}},
\end{equation}
we obtain for the mass flow rate,
\begin{multline}
\Phi=\frac{\rho h \tG}{2\tm}\\
\bigg\{ f(\thsub)\left(\tfrac{2}{3}-\epsub-\tfrac{1}{3}\epsub^{3} + \left(1-[1-\epsub]^{2}\right) [1-\epsub]\right)   \\
+f(\thalign)\left([1-\epsub]^{2}[1-\epsub-\epmid]-\tfrac{1}{3}\left([1-\epsub]^{3}-\epmid^{3}\right)\right) \\
 +\tfrac{2}{3}f(0)\epmid^{3}  \bigg\}.  \label{eqn:HighFlowFlux}
\end{multline}

\begin{figure}
\centering
\includegraphics[width=80mm]{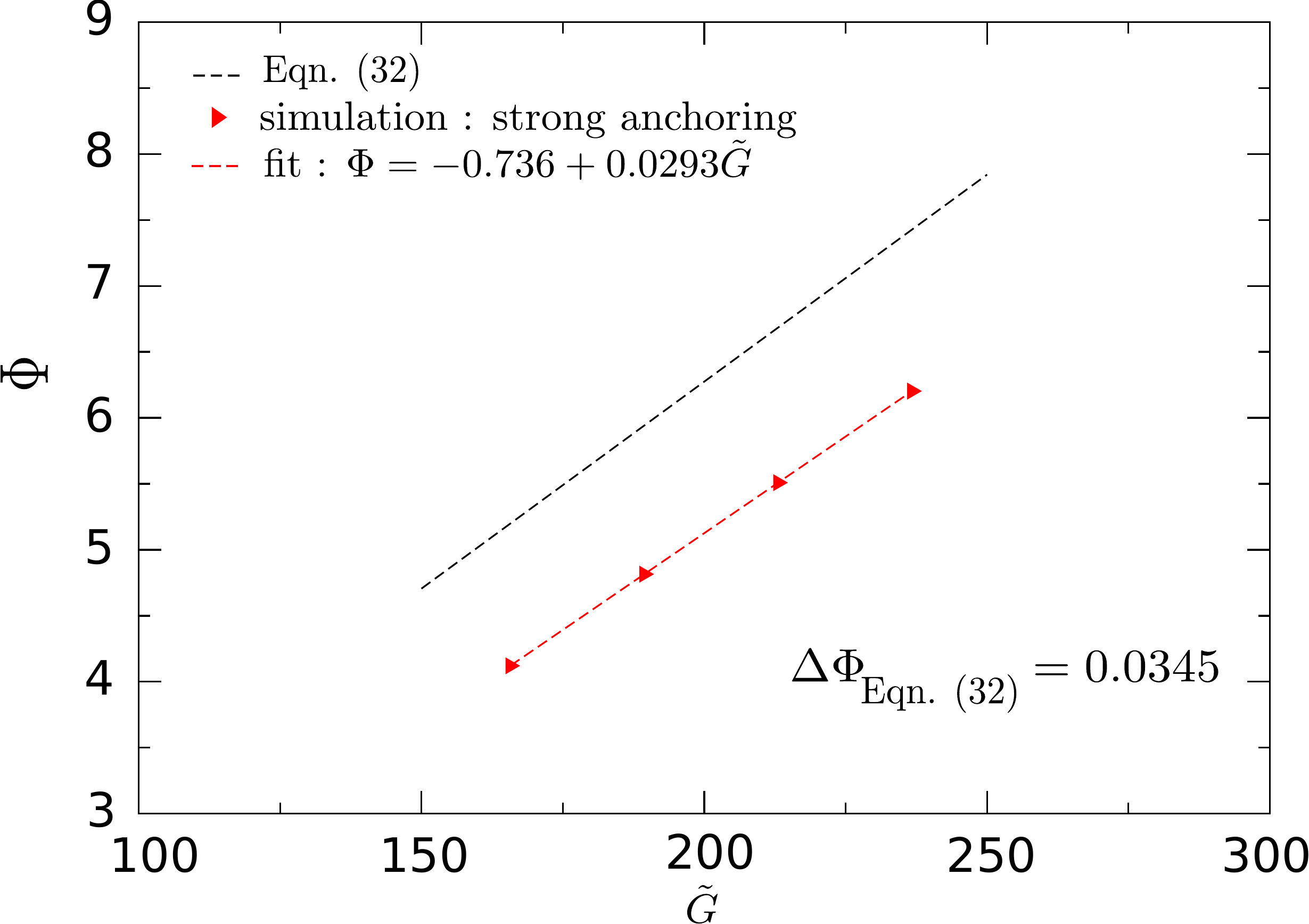}
\caption{Mass flow rate $\Phi$ as a function of the $\tG$ via simulation for weak and strong anchoring. Symbols represent simulation results, the dashed red line represents 
a linear fit to the simulation data and the black dashed line is the flux as obtained from Eqn.~(\ref{eqn:HighFlowFlux}).}
\label{fig:HighFlowFlux}
\end{figure}

The comparison of the flux and also the gradient $\dd\Phi/\dd\tG$ obtained via simulation with that given by Eqn.~(\ref{eqn:HighFlowFlux}) is 
shown in figure~(\ref{fig:HighFlowFlux}). In this figure, the mass flux obtained via simulation is shown by filled red symbols (for a system of $h=24$ where 
$\epsub=0.25$, $\epmid=0.20$ and $\thsub=46.6$ degrees, as can be seen in Fig.~(\ref{fig:HighFluxTheta})) whilst that obtained by Eqn.~(\ref{eqn:HighFlowFlux}) 
is represented by a black dashed line. Also given in this figure is a linear fit to simulation data shown by the red dashed line connecting the full triangular symbols. 
With this linear fit we are able to compare $\dd\Phi/\dd\tG$ obtained from simulation and that given by Eqn.~(\ref{eqn:HighFlowFlux}). Although $\dd\Phi/\dd\tG$ of 
Eqn.~(\ref{eqn:HighFlowFlux}) shows very good agreement with the simulations ($0.0345$ from Eqn.~(\ref{eqn:HighFlowFlux}) and $0.0293$ from the fit to simulation data), 
we note a discrepancy in the actual values of $\Phi$ due to the simulations results having a non-zero intercept. In order to resolve this discrepancy, 
a fuller treatment that goes beyond linear terms is required. A possible method to do this would be similar to what is done by Atkin and Leslie 
for Couette flow of nematic liquid crystals \cite{atkin70}, namely to produce a general solution in integral terms. In our case, the driving force 
produces a factor of $z$ in (17), which makes the integration more complicated, 
but in principle it should be possible.


\subsection{Planar and hybrid anchoring conditions}
\label{sec:planar}

\begin{figure*}
\centering
\includegraphics[width=115mm]{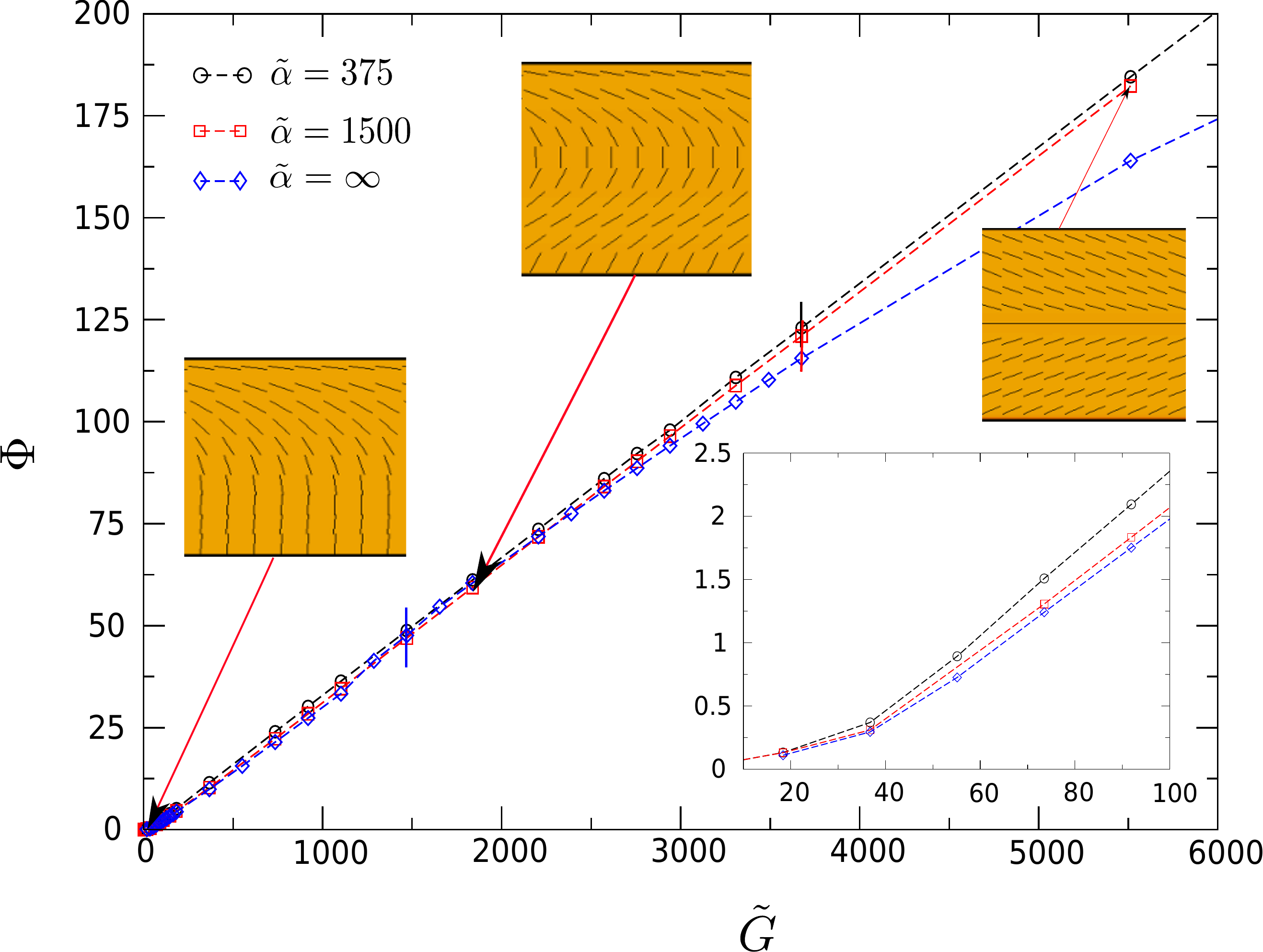}
\caption{Mass flow rate $\Phi$ as a function of the pressure gradient via simulation for (weak $\ta=375$, $\ta=1500$ to strong $\ta=\infty$) hybrid anchoring. 
The director orientation (black lines) obtained for different pressures gradients are shown in the 
embedded figures for the case of the 'vertical' state [at low $\tG$ (lower left) and near the jump in flow rate (top centre)] 
and 'horizontal' state (central right). Vertical lines represent the pressure gradient above which we observe the 'horizontal' state. 
The embedded graph is a blow-up of the graph for low pressure gradients.}
\label{fig:HybridFluxAlpha}
\end{figure*}

In this section we extend our work beyond homeotropic anchoring and perform lattice Boltzmann simulations of driven flow between two parallel 
plates for the cases where the anchoring at both walls of the channel is planar and where one wall has homeotropic anchoring and the other planar (which we shall term hybrid anchoring). 
In general, planar anchoring may be degenerate (favouring any orientation within the plane of the substrate equally) or non-degenerate 
(favouring one given direction within this plane). Here we restrict our study to the case of non-degenerate planar anchoring along the channel 
direction $x$. We set $l=400$ and $h=99$ and, as was 
previously done for the homeotropic anchoring study, we evolve the system until a steady state is obtained. Depending on the 
driving $\tG$, simulations were run $1\times10^6$ times steps (or $3$ to $4$ times longer if needed).

First we consider the case of hybrid anchoring where the substrate at $\tz=-1$ has homeotropic anchoring and that at $\tz=1$ has planar anchoring. 
In figure~(\ref{fig:HybridFluxAlpha}) we plot mass flow rate $\Phi$ as a function of $\tG$ for hybrid anchoring condition for weak $\ta=375$, $\ta=1500$ 
to strong $\ta=\infty$ anchoring strengths. Symbols represent simulation data and discontinuous lines are added to help guide the eye. The insets show the nematic 
configuration recorded for the indicated data points for the hybrid case. As before, the orange shading represents the nematic order parameter $S$ and the 
black lines depict the director $\mathbf{n}$. Embedded graph is a blow-up of the graph for low pressure gradients, which illustrates the 
jump in mass flow as the pressure gradient is increased, as previously seen for the homeotropic case. 
The lower-right inset represents the texture for low flow. In this case, the director orientation near the walls 
suffers only small distortions from the imposed anchoring conditions (i.e. $\theta\sim 90^{\circ}$ at $\tz=-1$ and $\theta\sim 180^{\circ}$ at $\tz=1$). 
Furthermore, the behaviour of $\theta$ as a function of $\tz$ is shown in figure~(\ref{fig:HybridTheta}), where the black continuous line illustrates 
this gradual increase of $\theta$ throughout the channel, from $\sim 90^{\circ}$ at $\tz=-1$ to $180^{\circ}$ at the other wall at $\tz=1$. 
Note that we are only considering one of two possible starting configurations relative to the flow; the mirror image, $\sim 90^{\circ}$ at 
$\tz=-1$ to $0^{\circ}$ at $\tz=1$, is equivalent in the case of zero flow and could equally well be the starting point, but we do not 
consider this case here.

\begin{figure}
\centering
\includegraphics[width=80mm]{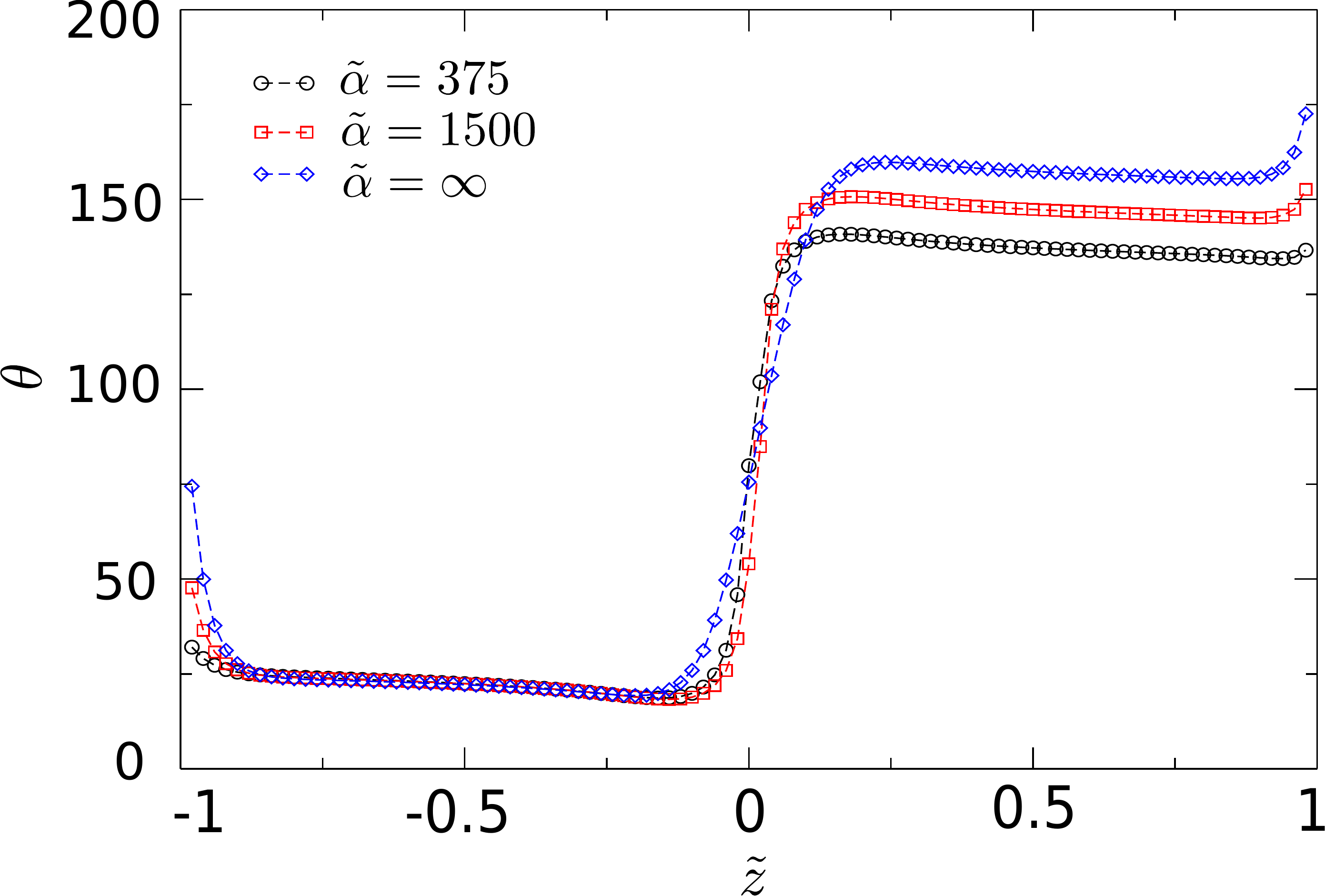}
\caption{Director profile as a function of channel height $\tz$ near the $\mathbb{V}\rightarrow\mathbb{H}$ transition. 
Different curves correspond to distinct values of $\ta$ for the hybrid anchoring condition.}
\label{fig:theta_nearVH}
\end{figure}

\begin{figure}
\centering
\includegraphics[width=80mm]{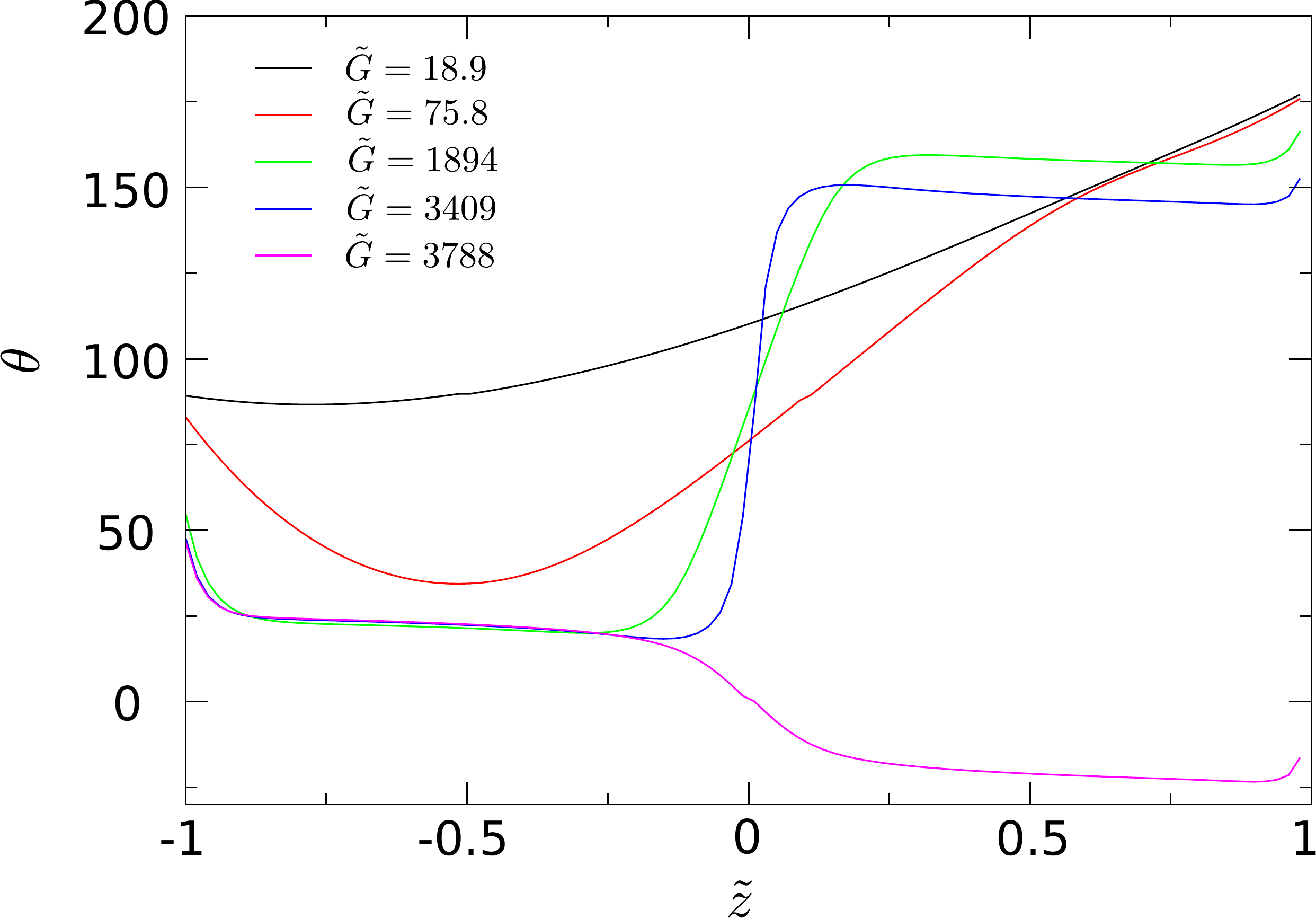}
\caption{Director angle $\theta$ across the width of the channel with hybrid anchoring of strength $\ta=1500$.}
\label{fig:HybridTheta}
\end{figure}

As $\tG$ is increased, the director profile takes a `question mark' shape as shown in the middle inset of 
figure~(\ref{fig:HybridFluxAlpha}) which resembles the behaviour previously reported for 
strong anchoring conditions \cite{marenduzzo04}. Similar to what was observed for the homeotropic case, the pressure gradient for which we 
observe the jump of mass flux increases with anchoring strength since the director profile will distort with more ease if anchoring is weaker. 
During the jump of flow rate in $\mathbb{V}$, the vertical state, there is significant change in $\theta$ across the entire width of the channel which 
can be seen in the behaviour of $\theta$ throughout the channel going from the red to green curves of figure~(\ref{fig:HybridFluxAlpha}). As in the 
homeotropic anchoring case, as $\tG$ is increased further, the director profile undergoes an abrupt transition so that its orientation at the 
middle of the channel becomes horizontal. We see from figure~(\ref{fig:HybridTheta}), that in this transition $\mathbb{V}\rightarrow\mathbb{H}$ 
(from the blue to pink curves), there is little change in the director profile near the walls, and change is limited towards the centre of the channel. 
As in the homeotropic case, we see that weak anchoring impedes rather than assists the $\mathbb{V}\rightarrow\mathbb{H}$ transition. However, 
contrary to the homeotropic case and considering the same anchoring strengths, we suggest from the analysis of figure (\ref{fig:theta_nearVH}), 
that the anchoring strength may play a larger role in the regions near the centre and top half of the channel.

Finally, we consider the case of planar anchoring on both walls, the behaviour of which turns out to be less rich than the homeotropic and hybrid substrates. 
In this case, the director is always in the $\mathbb{H}$ configuration. As the flow increases, the director at the walls becomes increasingly inclined away 
from the horizontal. In figure~(\ref{fig:HybridTheta}), the relation between $\Phi$ and $\tG$ shows only mild non-linearity, and does not exhibit the abrupt jump that 
was seen in the homeotropic case. This is in agreement with experiments carried out on cylindrical~\citep{fishers69} and rectangular~\citep{sengupta13IJMS} channels, 
which show less-pronounced departures from linearity for planar anchoring compared to homeotropic anchoring.

\begin{figure}
\centering
\includegraphics[width=80mm]{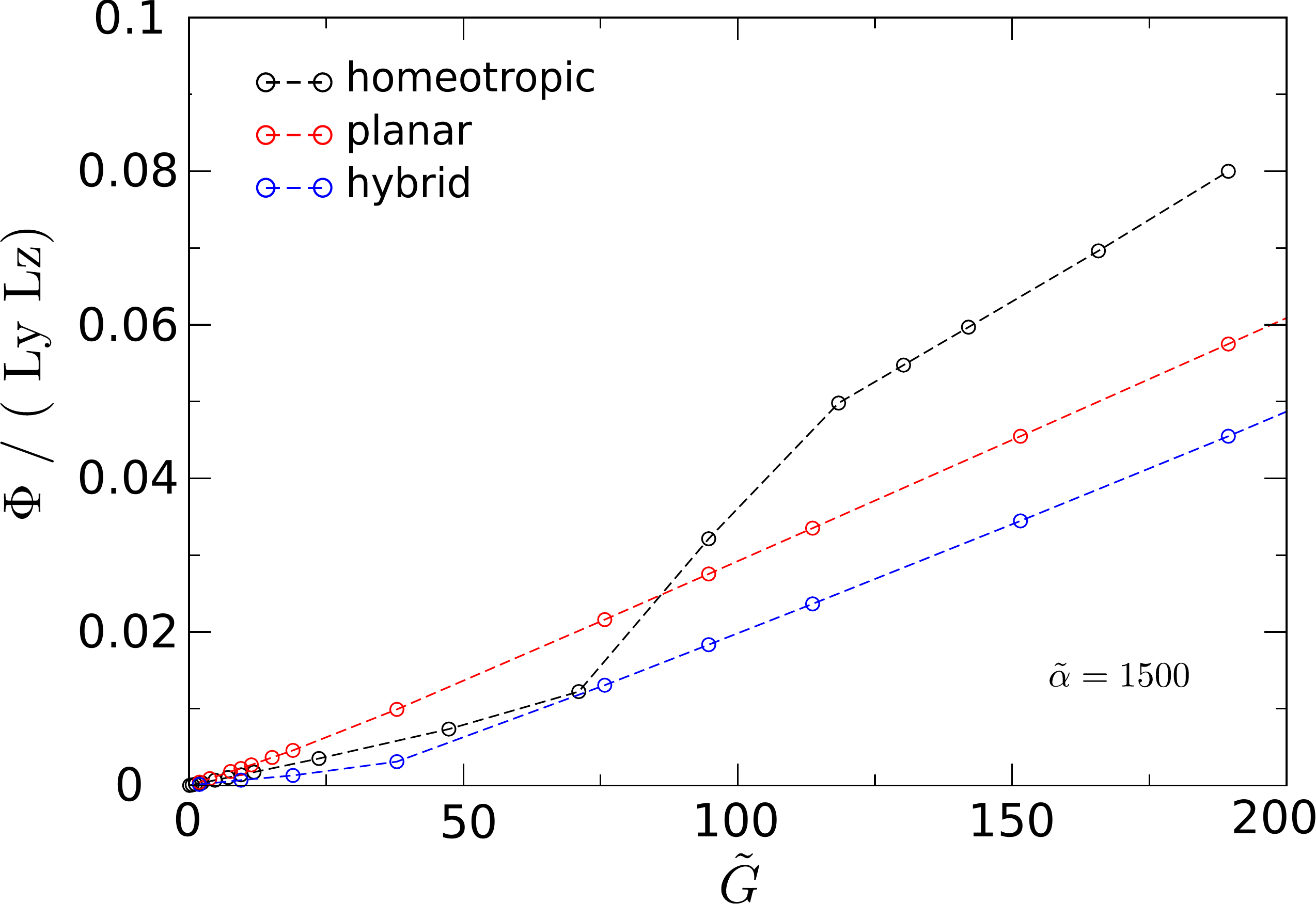}
\caption{Mass flow rate $\Phi$ per node as a function of the pressure gradient via simulation for homeotropic, planar and hybrid anchoring with $\ta=1500$.}
\label{fig:CompareFluxAnchoring}
\end{figure}

We briefly consider how the low flow limit in the planar case differs from that of the homeotropic case. Since the unperturbed $\theta$ is $0$ instead of $\pi/2$, 
the first-order expansions of Eqns.~(\ref{eqn:velocityODE},\ref{eqn:angleODE},\ref{eqn:subtrateEquilibriumDimless}) become
\begin{align}
\tm\left(1+r\left(\zeta-1\right)^{2}\right)\frac{\dd u_{1}}{\dd\tz}&=-\tz,  \\
\frac{\dd^{2}\theta_{1}}{\dd\tz^{2}}&= \frac{r\left(\zeta-1\right)}{4\left(1-r\left(\zeta-1\right)^{2}\right)}\tz, \\
\frac{\dd\theta_{1}}{\dd\tz}&=\pm\ta\theta_{1}\;\text{at}\;\tz=\pm 1.
\end{align}
{\em i.e.}, in eqns.~(\ref{eqn:velocityODE},\ref{eqn:angleODE}) $\zeta+1$ is replaced by $\zeta-1$. We thus derive
\begin{align}
u_{1}&=\frac{1}{2\tm\left(1+r\left(\zeta-1\right)^{2}\right)}\left(1-\tz^{2}\right),  \\
\theta_{1}&=-\frac{r\left(\zeta-1\right)}{24\left(1+r\left(\zeta-1\right)^{2}\right)}\left\{\left(\frac{\ta+3}{\ta+1}\right)\tz-\tz^{3}\right\}, \\
\Phi&=\rho h \frac{1}{3\tm\left(1+r\left(\zeta-1\right)^{2}\right)}\tG +.... \label{eqn:fluxPlanar}
\end{align}
Comparison of Eqn.~(\ref{eqn:fluxPlanar}) with Eqn.~(\ref{LowFlowFlux}) shows that the low flow limit $d\Phi/d\tG$ is larger in the planar case, as is 
confirmed in figure~(\ref{fig:CompareFluxAnchoring}). Hence we note that the linear term of $\Phi$ is dependent on the {\em type} of anchoring, while 
dependence on the {\em strength} of anchoring only arrives in the cubic term.

As a comparison of the three types of anchoring considered, we plot, in figure~(\ref{fig:CompareFluxAnchoring}), mass flow rate $\Phi$ as a function 
of $\tG$ for the three distinct anchoring conditions considered, for a chosen anchoring strength of $\ta=1500$, a relatively strong anchoring. In summary, 
the behaviour of flux as a function of $\tG$ for hybrid anchoring is reminiscent of what is observed in the homeotropic case with the exception that the 
height of the jump observed in the vertical state is somewhat smaller, being lowered by the presence of the wall with planar anchoring conditions. 
The director profile (and $\theta$) for sufficiently high $\tG$ in the $\mathbb{H}$ state is also similar to what is observed in the homeotropic 
anchoring case. In contrast to what is observed in the $\mathbb{H}$ state and as expected given the different anchoring conditions on the walls, 
the director profile observed in $\mathbb{H}$ state differs from other anchoring cases.

\section{Conclusions}
\label{sec:conclusions}
We have performed lattice Boltzmann simulations of driven flow of a nematic liquid crystal in two-dimensional channels, 
recording the mass flow rate $\Phi$ as a function of the applied pressure gradient $\tG$ for a variety of types and strengths of anchoring at 
the channel walls. Our results for homeotropic anchoring are summarised in figure~(\ref{fig:FluxAlpha}). At lower values of $\tG$, the nematic 
adopts a morphology $\mathbb{V}$ where the director is oriented to the vertical in the centre of the channel. We calculated the low flow limit 
from the dynamical equations and verified this against the results from the simulations. As $\tG$ increases, $d\Phi/d\tG$ becomes steeper as 
flow-induced distortions in the director lead to a reduction in the effective viscosity. $\Phi$ undergoes an abrupt jump, before moderating to an approximately 
linear rise with higher gradient than before the jump. At higher $\tG$, the nematic undergoes a topological transition to a 
state $\mathbb{H}$, where the director is orientated horizontal at the centre of the channel. We showed that this transition is driven by a dynamical instability due to 
the director being perturbed beyond the Leslie angle.

Figure~(\ref{fig:CompareFluxAnchoring}) compares results for homeotropic, planar and hybrid anchoring. We find that the curve for hybrid anchoring exhibits a smaller, 
but still notable jump, and there is a similar transition to the $\mathbb{H}$ state. For the case of planar anchoring, there are no striking departures from linearity, 
and the system is in the $\mathbb{H}$ state throughout.

The results of this paper demonstrate that the type and strength of anchoring at the channel has a profound effect on the relation between the 
mass flow rate $\Phi$ and the applied pressure gradient $\tG$ at the walls of the channel. Understanding these effects may help to fine-tune the 
flow properties of microfluidic nematic devices.   Conversely, the $\Phi$,$\tG$ curve provides a fingerprint to the anchoring that exists in the channel. 
This fingerprint can be identified using both qualitative features, such as abrupt departures from linearity, and quantitative features, such as the slope and 
subsequent derivatives of the curve.

An obvious direction for future work is to extend our simulations to three-dimensional channel flow, such as flow in a cylindrical or rectangular channel. 
We expect that frustration between competing boundary conditions and the formation of topological defects in the bulk will make flow behaviour in such geometries 
considerably more complicated than the two-dimensional system studied here, but we may hope that some of the principles we elucidated for two dimensions will carry forward 
into three. Experimental measurements on rectangular channels has been carried out by~\citet{sengupta13SM}. Figure~2 of Sengupta's paper appears to show a 
sharp decrease in the effective viscosity ($\propto \tG/\Phi$) at low driving pressures for homeotropic anchoring, reminiscent of the sharp increase we observe in our simulations.

\section*{Acknowledgements}

We thank Anupam Sengupta and Nuno Silvestre for fruitful discussions. We acknowledge the support of the Portuguese Foundation for Science and 
Technology (FCT) through the grants SFRH/BPD/73028/2010 (MLB), and PEst-OE/FIS/UI0618/2014 and EXCL/FIS-NAN/0083/2012 (all of us).

\section*{\appendix\appendixname}

In this appendix, we present a brief summary of the test performed on the hybrid lattice Boltzmann method applied throughout this work. 
We consider a (flow aligning) liquid crystal which is confined between two walls, both parallel to the $xz$-plane, distance $h$ apart. 
The lower one is at rest while the upper one is moved in the $x$-direction with velocity $v_0$. In this simple test case we consider 
strong homeotropic anchoring conditions and choose two test systems: one with $h=24$ and another where $h=99$, setting 
$v_0=0.01$ and $v_0=0.0025$, respectively. Shearing in the $x$-direction will cause the director to vary as a function 
of $\tz$ (i.e. $\theta=\theta(\tz)$). 
For high enough shear, $\theta$ reaches a saturation value of $\thalign$ in the bulk. $\theta$ varies continuously 
from the value defined by the boundary conditions to the bulk value. This behaviour has been both verified numerically and by 
theory (e.g. in reference~\citenum{degennes93} and \citenum{Carlsson84} and references therein). 

\begin{figure}
\centering
\includegraphics[width=80mm]{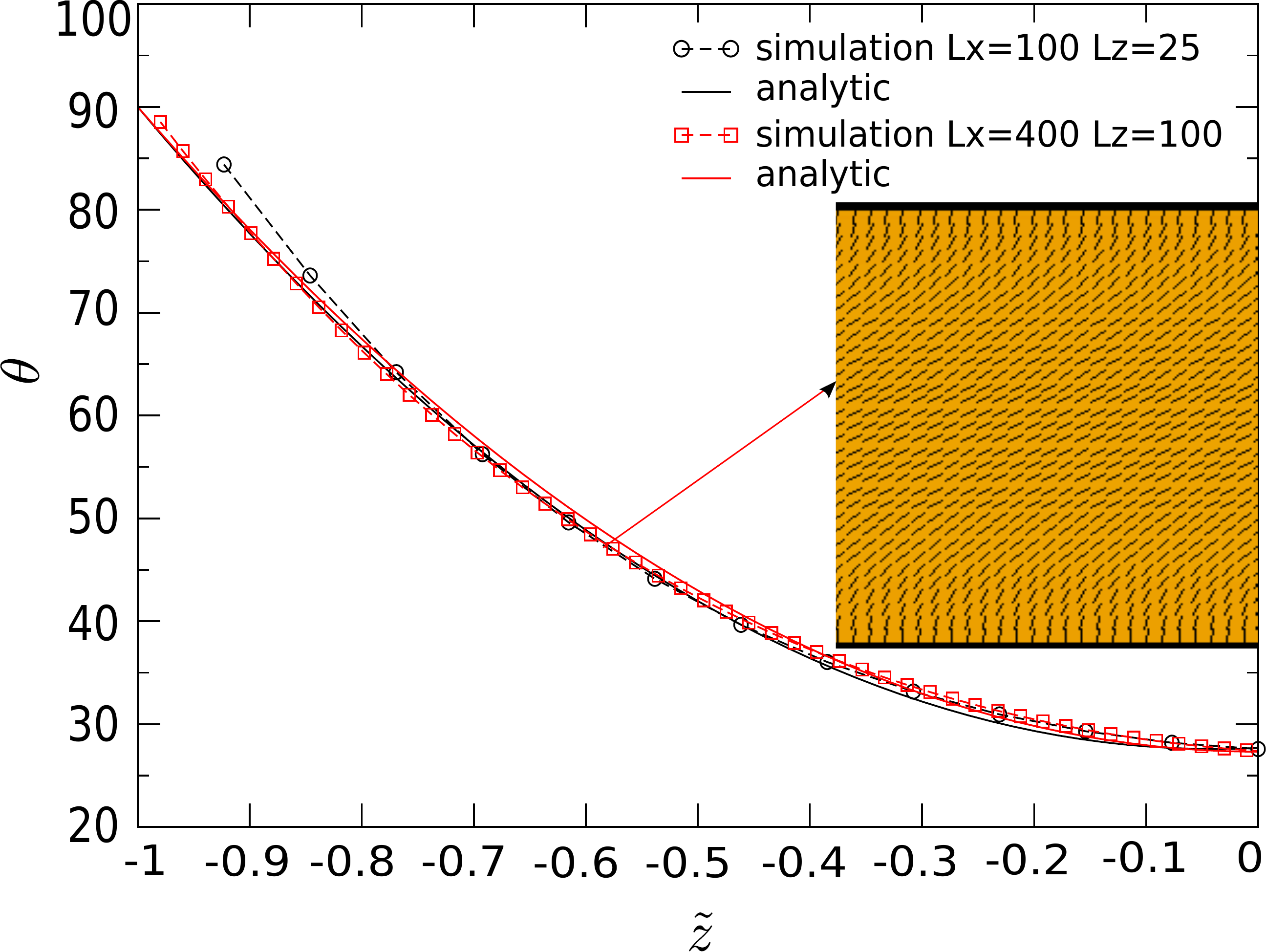}
\caption{Director profile for simple shear of a nematic liquid crystal between two parallel walls. Lower wall is kept at rest while the top one is sheared with 
velocity $v_0$. Continuous lines represent the analytic solution whereas symbols connected by discontinuous lines are numerical results. 
Inset: Example of the director orientation (black lines) obtained.}
\label{fig:Sheartheta}
\end{figure}

Figure~(\ref{fig:Sheartheta}) compares the director profile $\theta(\tz)$ obtained numerically in this work and the 
analytic solution (Eqn. III$.9$ of reference \citenum{Carlsson84}) for the two systems studied. 
The inset of this figure illustrates the trend of the director in the sheared channels. 
We observe that numerical results for the larger system are in very good agreement with theory differing by $\lesssim 1\%$. The smaller system 
also presents good agreement between numerical results and theory but in this case they mostly differ near the walls by $\lesssim 6\%$. In both systems, 
the director at mid-channel presents $\theta(\tz=0)=27$.

\footnotesize{
\bibliography{paper} 
\bibliographystyle{rsc} 
}

\end{document}